\documentclass[traditabstract]{aa}  

\usepackage{graphicx}

\usepackage{txfonts}
\usepackage{natbib}
\begin{document}

   \title{Absorbed relativistic jets in radio-quiet narrow-line Seyfert 1 galaxies}

%   \subtitle{}

   \author{M. Berton\inst{1,2}\thanks{e-mail: marco.berton@utu.fi}
	\and E. J\"arvel\"a\inst{3}\thanks{e-mail: ejarvela@sciops.esa.int}
	\and L. Crepaldi\inst{4} 
	\and \\ A. L\"ahteenm\"aki\inst{2,5}
	\and M. Tornikoski\inst{2} 
	\and E. Congiu\inst{6}
	\and P. Kharb\inst{7}
	\and G. Terreran\inst{8}
	\and A. Vietri\inst{4}
          }

   \institute{
	$^1$ Finnish Centre for Astronomy with ESO (FINCA), University of Turku, Vesilinnantie 5, FI-20014 University of Turku, Finland; \\
	$^2$ Aalto University Mets{\"a}hovi Radio Observatory, Mets{\"a}hovintie 114, FI-02540 Kylm{\"a}l{\"a}, Finland; \\
	$^3$ European Space Agency, European Space Astronomy Centre, C/ Bajo el Castillo s/n, 28692 Villanueva de la Ca\~nada, Madrid, Spain; \\
	$^4$ Dipartimento di Fisica e Astronomia "G. Galilei", Universit\`a di Padova, Vicolo dell'Osservatorio 3, 35122 Padova, Italy; \\
	$^5$ Aalto University Department of Electronics and Nanoengineering, P.O. Box 15500, FI-00076, Aalto, Finland; \\
	$^6$ Las Campanas Observatory, Carnegie Institution of Washington, Colina El Pino, Casilla 601, La Serena, Chile; \\
	$^7$ National Centre for Radio Astrophysics - Tata Institute of Fundamental Research, Post Bag 3, Ganeshkhind, Pune 411007, India;\\
	$^8$ Center for Interdisciplinary Exploration and Research in Astrophysics CIERA, Department of Physics and Astronomy, Northwestern University, Evanston, IL 60208, USA. \\
             }

   \date{}

\authorrunning{M. Berton et al.}
\titlerunning{Absorbed relativistic jets in NLS1s}
 
  \abstract{Narrow-line Seyfert 1 (NLS1) galaxies are peculiar active galactic nuclei (AGN). Most of them do not show strong radio emission, but recently seven radio-quiet (or -silent) NLS1s have been detected flaring multiple times at 37~GHz by the Mets\"ahovi Radio Telescope, indicating the presence of relativistic jets in these peculiar sources. We observed them with the Karl G. Jansky Very Large Array (JVLA) in A configuration at 1.6, 5.2, and 9.0 GHz. Our results show that these sources are either extremely faint or not detected in the JVLA bands. At those frequencies, the radio emission from their relativistic jet must be absorbed, either via synchrotron self-absorption as it occurs in gigahertz-peaked sources or, more likely, via free-free absorption by a screen of ionized gas associated with starburst activity or shocks. Our findings cast new shadows on the radio-loudness criterion, which seems to be more and more frequently a misleading parameter. New high-frequency and high-resolution radio observations are essential to test our hypotheses.}


   \maketitle
%
%-------------------------------------------------------------------
\newcommand{\kms}{km s$^{-1}$}
\newcommand{\ergs}{erg s$^{-1}$}
\section{Introduction}
In approximately 10\% of active galactic nuclei (AGN) the system black hole/accretion disk is capable of launching powerful relativistic plasma jets, which in some cases have enough energy to propagate outside the host galaxy. 
Traditionally, the signature of relativistic jets is strong radio emission due to synchrotron radiation. 
Conversely, the relativistic jet is typically less dominant in the optical band.  
For these reasons, the radio-loudness criterion introduced by \citet{Kellermann89}, defined as the ratio between the flux density at 5 GHz and the optical B-band flux density, is usually good at separating sources with and without relativistic jets. 
When the ratio is larger than 10, the source is classified as radio-loud. 
Conversely, it is radio-quiet. 
The power and luminosity of relativistic jets depend on the black hole mass and follow a non-linear scaling relation. 
Therefore, low-mass AGN will inevitably harbor less powerful jets \citep{Heinz03, Foschini14}. 
Since the scaling weighs more on the radio luminosity than in the optical, the radio-loudness criterion can become inadequate in low-mass sources. \par
\begin{table*}[!t]
\caption{List of radio-quiet/silent NLS1s detected at 37~GHz.}
\label{tab:source_list}
\centering
\footnotesize
\begin{tabular}{lcccccccccc }
\hline\hline
SDSS Name & Short alias & z & Scale & $S^{\rm ave}_{37 \; \rm{GHz}}$ & $S^{\rm max}_{37 \; \rm{GHz}}$ & $N_{\rm det}$ & JVLA Date & Time (L) & Time (C)& Time (X) \\
\hline\hline
SDSS J102906.69+555625.2 & J1029+5556 & 0.451 & 9.685 & 0.42 & 0.52 & 3 & 2019-09-08 & 1012 & 714 & 714 \\
SDSS J122844.81+501751.2 & J1228+5017 & 0.262 & 5.657 & 0.41 & 0.51 & 5 & 2019-09-22 & 1070 & 772 & 774 \\
SDSS J123220.11+495721.8 & J1232+4957 & 0.262 & 5.626 & 0.46 & 0.56 & 4 & 2019-09-27 & 1072 & 738 & 772 \\
SDSS J150916.18+613716.7 & J1509+6137 & 0.201 & 4.313 & 0.67 & 0.97 & 13 & 2019-09-26 & 1070 & 772 & 772 \\
SDSS J151020.06+554722.0 & J1510+5547 & 0.150 & 3.214 & 0.45 & 0.83 & 15 & 2019-09-23 & 1070 & 772 & 770 \\
SDSS J152205.41+393441.3 & J1522+3934 & 0.077 & 1.650 & 0.59 & 1.43 & 4 & 2019-09-24 & 1072 & 774 & 772 \\
SDSS J164100.10+345452.7 & J1641+3454 & 0.164 & 3.518 & 0.37 & 0.46 & 2 & 2019-09-22 & 1068 & 770 & 772 \\
 \hline 
\end{tabular}
\tablefoot{Columns: (1) source name in the SDSS; (2) short name; (3) redshift; (4) scale (kpc arcsec$^{-1}$); (5) average flux density measured at Mets\"ahovi at 37~GHz (in Jy); (6) maximum flux density measured at Mets\"ahovi at 37~GHz (in Jy); (7) number of detections at Mets\"ahovi; (8) date of JVLA observations; (9) exposure time at 1.6~GHz (L band) (seconds); (10) exposure time at 5.2~GHz (C band) (seconds); (11) exposure time at 9.0~GHz (X band) (seconds).}
\end{table*}
Narrow-line Seyfert 1 (NLS1) galaxies are a well-known example of low-mass AGN \citep{Peterson11}.  
Classified for the first time by \citet{Osterbrock85}, they are characterized by a low full width at half maximum (FWHM) of the H$\beta$ line, which is usually interpreted as low rotational velocity around a low-mass (10$^6$-10$^8$ M$_\odot$) black hole \citep[e.g.,][]{Rakshit17a, Chen18, Peterson18}. 
NLS1s belong to the so-called population A of the AGN main sequence \citep[MS,][]{Sulentic02, Sulentic15, Marziani18b}.
AGN indeed seem to live in a specific locus of the plane defined by the R4570 parameter, that is the relative intensity of Fe II multiplets with respect to H$\beta$, and by the FWHM(H$\beta$), also known as eigenvector 1 \citep{Boroson92}. 
The physical driver of the MS may be a combination of inclination and Eddington ratio \citep{Panda19}. 
The latter is very high in population A sources like NLS1s (low FWHM and high R4570), and along with the low black hole mass and several other physical properties, this characteristic may indicate that NLS1s and possibly population A sources in general may be one of the early phases of AGN life \citep{Mathur00, Sulentic00, Fraixburnet17b, Berton17, Berton20a}. \par
A small fraction ($\sim$7\%, \citealp{Komossa06}) of NLS1s can be classified as radio-loud and likely harbor relativistic jets. 
Their jet power is low when compared to the flat-spectrum radio quasars (FSRQs), but comparable to those of BL Lacertae objects (BL Lacs, \citealp{Foschini15}).
In BL Lacs, the low jet power can be attributed to an inefficient accretion mechanism \citep{Heckman14, Foschini17}. 
In FSRQs and NLS1s, instead, the accretion mechanism is the same and their jet power, as expected, follows the non-linear scaling with the black hole mass \citep{Foschini17}. \par
The vast majority of the NLS1 population is either radio-quiet or never detected at any radio frequency (radio-silent). 
The low radio luminosity compared to the optical luminosity may be due to the absence of relativistic jets. 
However, most radio-loud NLS1s are found in the high-mass part of the NLS1 distribution, while radio-quiet and radio-silent objects have lower black hole masses \citep[][albeit this could be a selection effect, see \citealp{Jarvela17}]{Foschini15, Berton15a, Cracco16}. 
If these low-mass objects had relativistic jets they would have a rather low radio luminosity due to the low jet power, which would lead to the radio-quiet classification.
Furthermore, even relatively powerful relativistic jets, when misaligned, could be faint at radio frequencies because we would not see any relativistic beaming effects \citep{Berton18a}. 
Therefore, sources with relativistic jets could be hidden among the radio-quiet/silent population \citep{Foschini11, Foschini12, Berton15a}. \par
The detection of radio-quiet and radio-silent NLS1s at 37~GHz with the Mets\"ahovi Radio Telescope (Finland) at flux densities close to 1 Jy, although unexpected, might be understood in this scenario \citep{Lahteenmaki17, Lahteenmaki18}. 
Albeit previously undetected in radio, some NLS1s harbor relativistic jets which can be seen even at high frequencies during flares. 
However, considering the nature of NLS1s, it is also possible that their relativistic jets are extremely young. 
Such sources have been observed in radio only by the FIRST and NVSS surveys \citep{Becker95, Condon98}. 
Their jets may have been launched at some point between the old observations and the Mets\"ahovi survey. 
Finally, one last possibility is that their radio spectra are extremely inverted ($\alpha \lesssim -2$, S$_\nu \propto \nu^{-\alpha}$) due to some form of absorption. 
Both free-free and synchrotron self-absorption could indeed absorb the radiation of the relativistic jet at low frequencies, while leaving high-frequency photons free to escape. \par
To discern between these diverse options, we observed seven of these sources with the Karl G. Jansky Very Large Array (JVLA). 
All of them have been detected multiple times at 37~GHz with the Mets\"ahovi Radio Telescope. 
In Sect.~2 we will describe the sample, in Sect.~3 the observations and the data analysis, in Sect.~4 we will present our results, in Sect.~5 we will discuss them, and in Sect.~6 we will provide a brief summary of our work. 
Throughout this paper, we adopt a standard $\rm \Lambda CDM$ cosmology, with a Hubble constant $H_0 = 70$ \kms\ Mpc$^{-1}$, and $\Omega_\Lambda = 0.73$ \citep{Komatsu11}. 
 
\section{Data reduction and analysis}

The sources were part of a sample of 66 objects selected according to their large-scale environment \citep{Jarvela17} and promising multiwavelength properties \citep{Jarvela15, Lahteenmaki18}. 
The host galaxy morphology has been studied only for a handful of them, but it is preferentially a spiral with ongoing interaction \citep[][J\"arvel\"a et al. in prep.]{Jarvela18}, a characteristic which seems common among jetted NLS1s \citep{Anton08, Kotilainen16, Olguiniglesias17, Berton19a, Olguiniglesias20}. 
Out of those 66, seven have been detected more than once at 37~GHz at Mets\"ahovi.
The basic properties of these sources are shown in Table~\ref{tab:source_list}. 
We observed these seven objects with the JVLA in A configuration in three different bands, L, C, and X, centered at 1.6 GHz, 5.2 GHz, and 9.0 GHz, and divided into 16 spectral windows. 
The total bandwidth was 1 GHz at 1.6~GHz, and 2 GHz at 5.2~GHz and 9.0~GHz.  
For each source we observed the NLS1 3C 286 \citep{Berton17} as main calibrator, and nearby bright unresolved sources for phase calibration. 
The project code is 19A-200 (P.I. Berton).
The expected noise levels were 20, 8, and 9 $\mu$Jy at 1.6~GHz, 5.2~GHz, and 9.0~GHz, respectively. \par
We reduced and analyzed the data using the Common Astronomy Software Applications (CASA) version 5.0.0-218 and the standard JVLA data reduction pipeline version 5.0.0. 
To produce the maps we used a pixel size of 0.15$^{\prime\prime}$ at 1.6~GHz, 0.05$^{\prime\prime}$ at 5.2~GHz, and 0.025$^{\prime\prime}$ at 9.0~GHz.
At all frequencies, first we examined a squared region of 2048 px centered on the source, that is large enough to cover the entire primary beam.  
If sidelobes of other nearby sources were clearly visible, we increased the size of the map in order to model also these neighboring sources. 
Otherwise, we just modeled the source using the CLEAN algorithm. 
In all cases but one, we reached a noise close to the theoretical level. 
We did not perform self-calibration on the data. \par

\begin{table}
\caption{Flux densities and luminosities in all bands.}
\label{tab:fluxes}
\centering
\scalebox{0.8}{
\footnotesize
\begin{tabular}{l c c c c c c c c c c c c c c} 
\hline
\textbf{1.6~GHz} \\
\hline
Name & noise & $S_{\rm p}$ & $S_{\rm i}$ & log($L_{\rm p}$) & log($L_{\rm i}$) \\
\hline
J1029+5556 & 26 & $<$0.081 & {} & $<$38.99 & {} \\ 
J1228+5017 & 23 & 0.685$\pm$0.024 & 0.899$\pm$0.067 & 39.35$\pm$0.04 & 39.47$\pm$0.07 \\ 
J1232+4957 & 21 & 0.093$\pm$0.021 & 0.234$\pm$0.041 & 38.44$\pm$0.23 & 38.84$\pm$0.18  \\ 
J1509+6137 & 22 & $<$0.069 & {} & $<$38.09 & {} \\ 
J1510+5547 & 21 & 0.074$\pm$0.022 & 0.245$\pm$0.039 & 37.84$\pm$0.30 & 38.36$\pm$0.16 \\ 
J1522+3934 & 21 & 1.030$\pm$0.021 & 2.030$\pm$0.160 & 38.38$\pm$0.02 & 38.67$\pm$0.08 \\ 
J1641+3454 & 27 & 1.721$\pm$0.028 & 2.520$\pm$0.097 & 39.31$\pm$0.02 & 39.48$\pm$0.04 \\ 
\hline\hline
\textbf{5.2~GHz} \\
\hline
Name & noise & $S_{\rm p}$ & $S_{\rm i}$ & log($L_{\rm p}$) & log($L_{\rm i}$)  \\
\hline
J1029+5556 & 7 & $<$0.024 & {} & $<$38.46 & {} \\ 
J1228+5017 & 8 & 0.248$\pm$0.009 & 0.304$\pm$0.013 & 38.91$\pm$0.04 & 39.00$\pm$0.04 \\ 
J1232+4957 & 7 & 0.057$\pm$0.008 & 0.050$\pm$0.008 & 38.23$\pm$0.14 & 38.17$\pm$0.16 \\ 
J1509+6137 & 8 & $<$0.027 & {} & $<$37.68 & {} \\ 
J1510+5547 & 8 & 0.032$\pm$0.009 & 0.052$\pm$0.017 & 37.48$\pm$0.28 & 37.69$\pm$0.33 \\ 
J1522+3934 & 12 & 0.329$\pm$0.013 & 0.380$\pm$0.039 & 37.88$\pm$0.04 & 37.95$\pm$0.1 \\ 
J1641+3454 & 8 & 0.518$\pm$0.009 & 0.698$\pm$0.035 & 38.79$\pm$0.02 & 38.92$\pm$0.05 \\ 
\hline\hline
\textbf{9.0~GHz} \\
\hline
Name & noise & $S_{\rm p}$ & $S_{\rm i}$ & log($L_{\rm p}$) & log($L_{\rm i}$) \\
\hline
J1029+5556 & 8 & $<$0.027 & {} & $<$38.31 & {} \\ 
J1228+5017 & 7 & 0.184$\pm$0.008 & 0.208$\pm$0.012 & 38.75$\pm$0.04 & 38.80$\pm$0.06 \\ 
J1232+4957 & 7 & $<$0.021 & {} & $<$37.97 & {} \\ 
J1509+6137 & 9 & $<$0.030 & {} & $<$37.65 & {} \\ 
J1510+5547 & 7 & 0.026$\pm$0.008 & 0.031$\pm$0.007 & 37.37$\pm$0.31 & 37.44$\pm$0.23 \\ 
J1522+3934 & 7 & 0.202$\pm$0.008 & 0.273$\pm$0.021 & 37.67$\pm$0.04 & 37.80$\pm$0.08 \\ 
J1641+3454 & 7 & 0.291$\pm$0.008 & 0.364$\pm$0.018 & 38.54$\pm$0.03 & 38.64$\pm$0.05 \\ 
\hline\hline
\end{tabular}
}
\tablefoot{Columns: (1) name; (2) image noise level (in $\mu$Jy); (3) peak flux (in mJy beam$^{-1}$); (4) integrated flux (in mJy); (5) logarithm of the peak luminosity (in \ergs); (6) logarithm of the integrated luminosity. }
\end{table}

\begin{table*}[!t]
\caption{Geometrical and physical properties of the sources.} 
\label{tab:measurements}
\centering
\scalebox{0.75}{
\footnotesize
\begin{tabular}{l c c c c c c c c c c c} 
\hline\hline
Name & Beam maj & Beam min & Beam PA & Core maj & Core min & Core PA & $\alpha_{LC}^p$ & $\alpha_{LC}^i$ & $\alpha_{CX}^p$ & $\alpha_{CX}^i$ &  log(T$_b$) \\
\hline
J1029+5556 & 0.47 & 0.41 & $+$37.29 & {} & {} & {} & {} & {} & {} &  {} &  {} \\
J1228+5017 & 0.49 & 0.40 & $+$74.84 & 0.230$\pm$0.045 & 0.188$\pm$0.068 & 124.0$\pm$48.0 & 0.86$\pm$0.06 & 0.92$\pm$0.10 & 0.54$\pm$0.15 & 0.69$\pm$0.18 & 2.68\\
J1232+4957 & 0.48 & 0.40 & $-$57.61 & 0.420$\pm$0.039 & 0.395$\pm$0.035 & 175.0$\pm$58.0 & 0.42$\pm$0.31 & 1.31$\pm$0.28 & $>$1.58 & $>$1.34 & 1.46\\
J1509+6137 & 0.50 & 0.39 & $-$37.97 & {} & {} & {} & {} & {} & {} &  {} &  {} \\
J1510+5547 & 0.45 & 0.40 & $-$14.58 & 0.462$\pm$0.210 & 0.202$\pm$0.180 & 17.0$\pm$24.0 & 0.71$\pm$0.49 & 1.32$\pm$0.41 & 0.38$\pm$1.07 & 0.94$\pm$1.01 & 1.42\\
J1522+3934 & 0.44 & 0.40 & $-$25.98 & 0.513$\pm$0.035 & 0.392$\pm$0.021 & 110.4$\pm$8.0 & 0.97$\pm$0.05 & 1.42$\pm$0.15 & 0.89$\pm$0.14 & 0.60$\pm$0.33 & 2.07\\
J1641+3454 & 0.42 & 0.41 & $-$44.75 & 0.320$\pm$0.031 & 0.152$\pm$0.045 & 163.6$\pm$8.0 & 1.02$\pm$0.03 & 1.09$\pm$0.08 & 1.05$\pm$0.08 & 1.19$\pm$0.18 & 2.92\\
\hline
\end{tabular}
}
\tablefoot{Columns: (1) name; (2) beam major axis (arcsec); (3) beam minor axis (arcsec); (4) beam position angle (degrees); (5) core major axis deconvolved from beam (arcsec); (6) core minor axis deconvolved from beam (arcsec); (7) core position angle (degrees); (8) spectral index of the peak flux density between 1.6~GHz and 5.2~GHz; (9) spectral index of the integrated flux density between 1.6~GHz and 5.2~GHz; (10) spectral index of the peak flux density between 5.2~GHz and 9.0~GHz; (11) spectral index of the integrated flux density between 5.2~GHz and 9.0~GHz; (12) logarithm of the brightness temperature.}
\end{table*}

\begin{figure*}[t]
    \centering
    \begin{minipage}[t]{.33\textwidth}
	\centering
	\includegraphics[trim={0cm 11cm 4cm 0cm}, width=\textwidth]{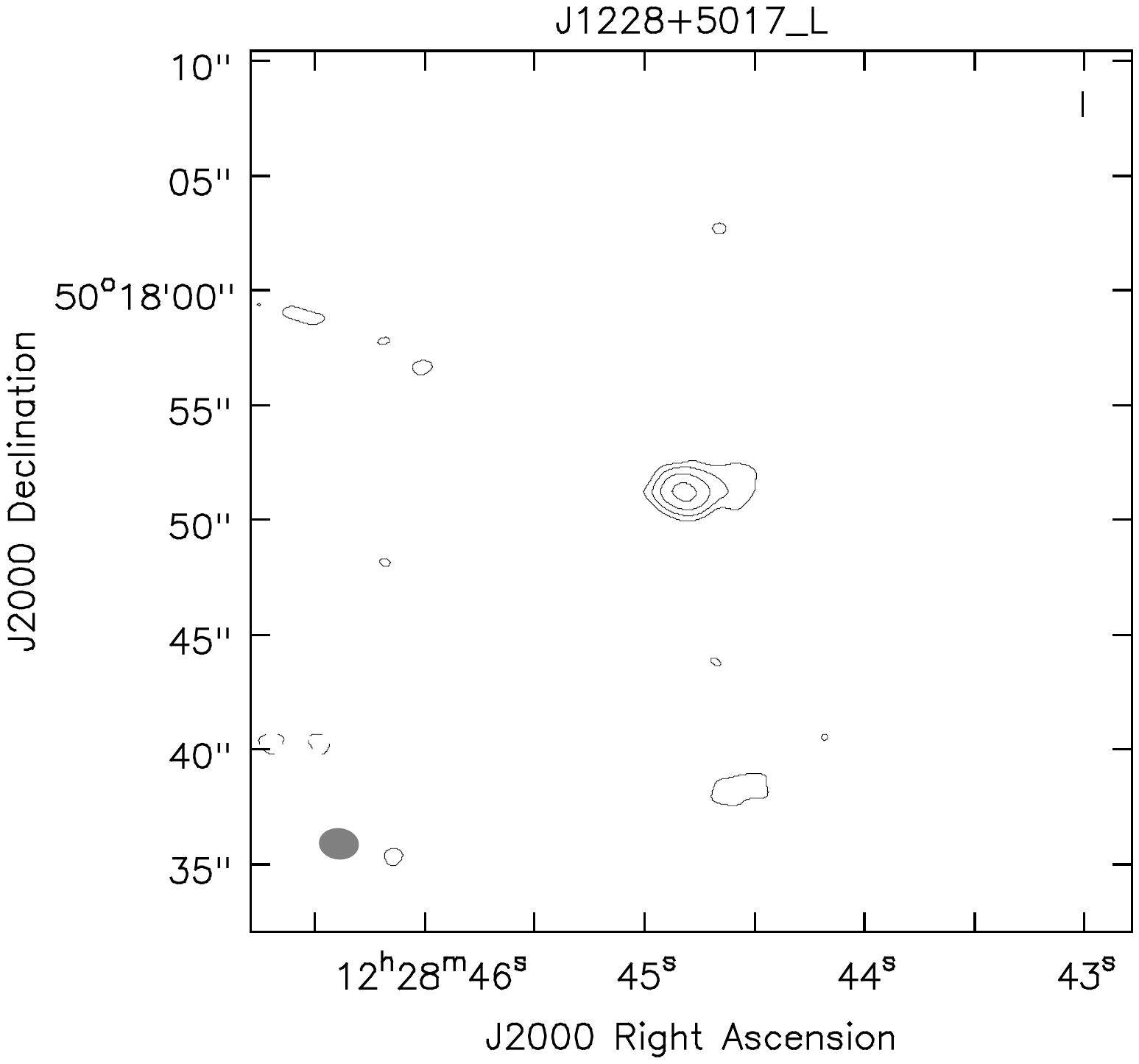}
    \end{minipage}
    \hfill
    \begin{minipage}[t]{.33\textwidth}
        \centering
	\includegraphics[trim={0cm 11cm 4cm 0cm}, width=\textwidth]{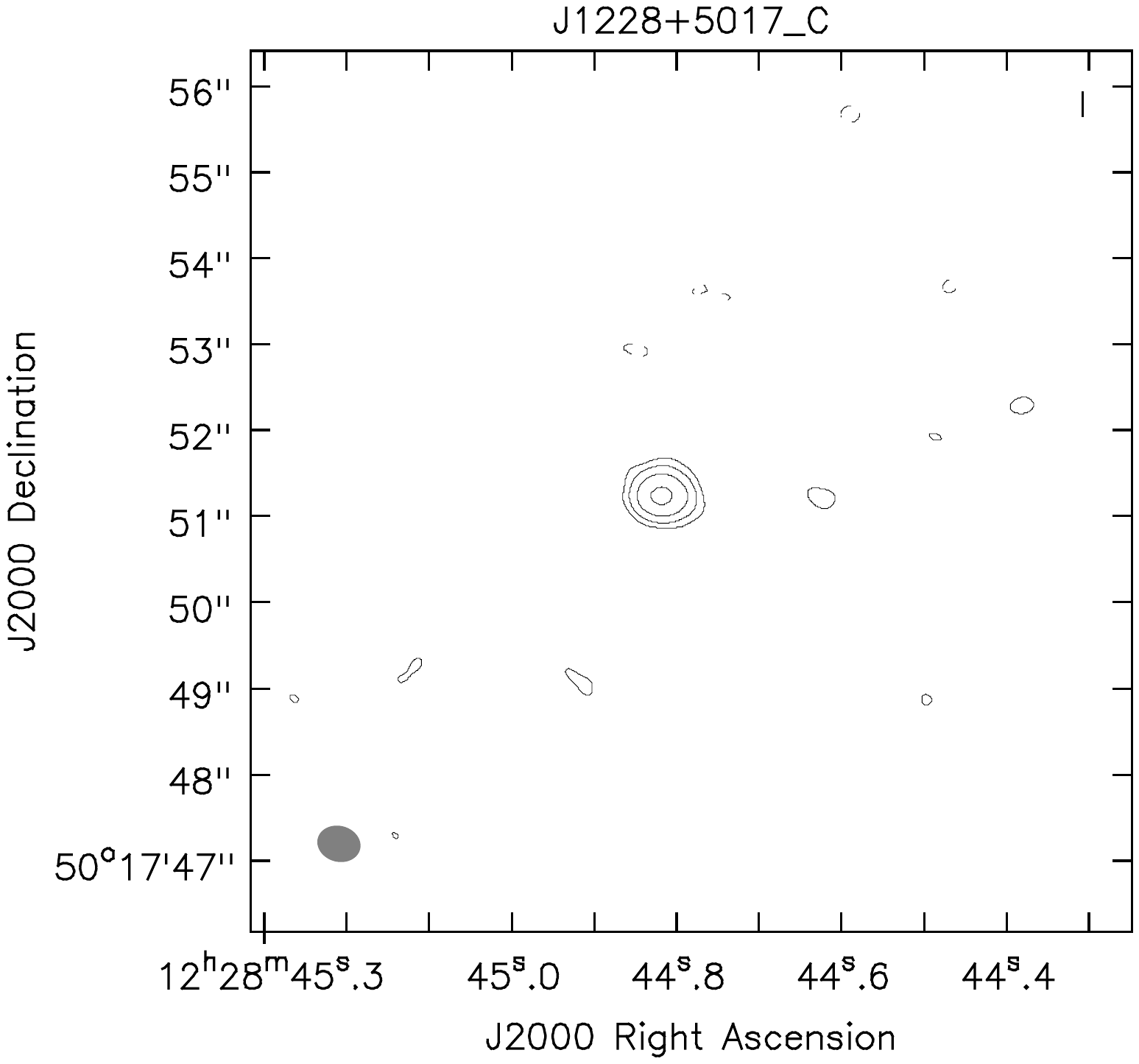}
    \end{minipage} 
    \hfill
    \begin{minipage}[t]{.33\textwidth}
        \centering
	\includegraphics[trim={0cm 11cm 4cm 0cm}, width=\textwidth]{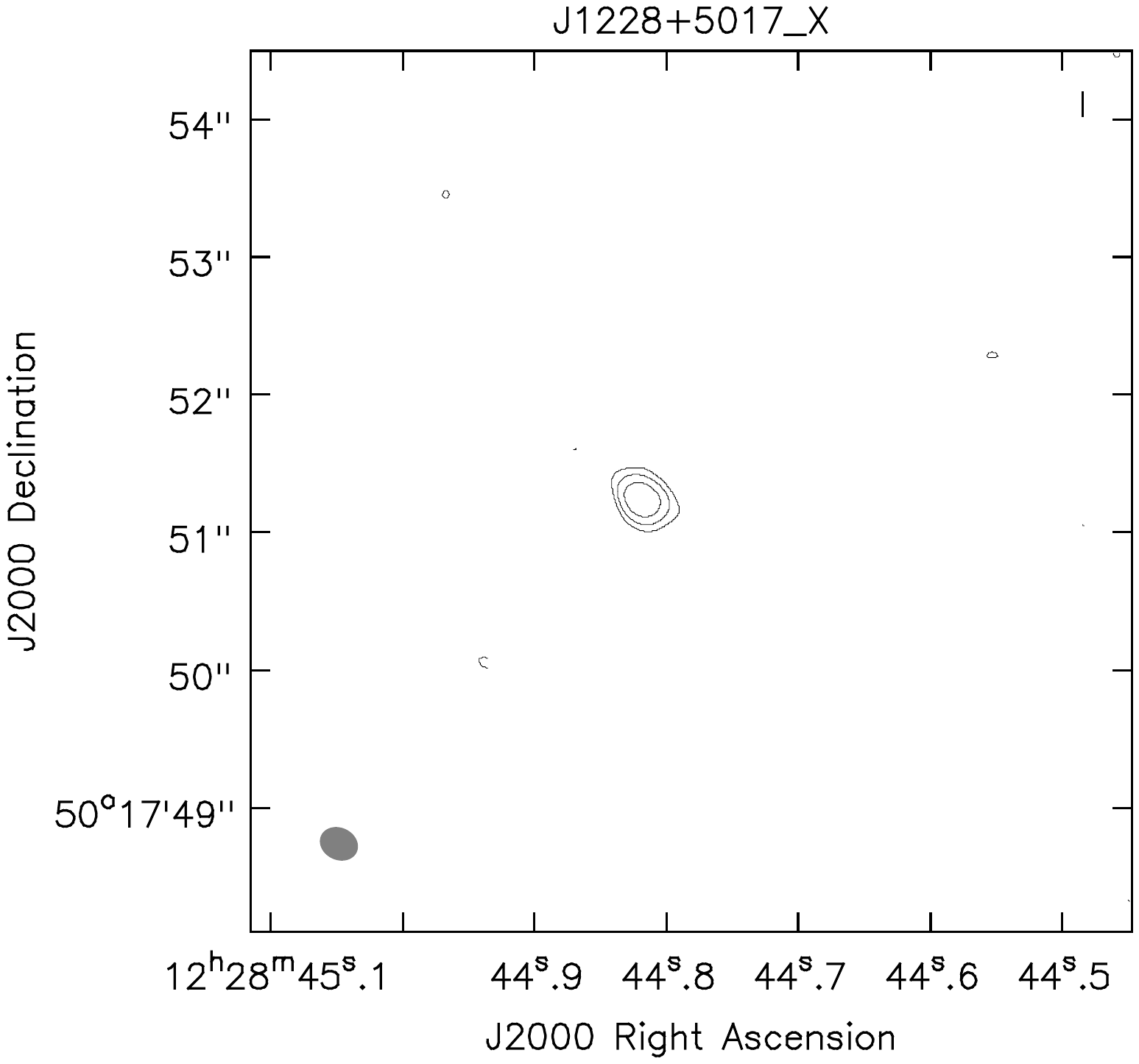}
    \end{minipage}  
\caption{From left to right, maps of J1228+5017 at 1.6 (L), 5.2 (C), and 9.0~GHz (X), respectively. The rms of the maps is 23, 8, and 7 $\mu$Jy, respectively, the contour levels are at rms$\times$(-3, 3$\times$2$^n$), n $\in$ [0,3], beam size 2.77$\times$2.32 kpc.}
\label{fig:J1228}
\end{figure*}

\begin{figure*}[t]
    \centering
    \begin{minipage}[t]{.33\textwidth}
	\centering
	\includegraphics[trim={0cm 11cm 4cm 0cm}, width=\textwidth]{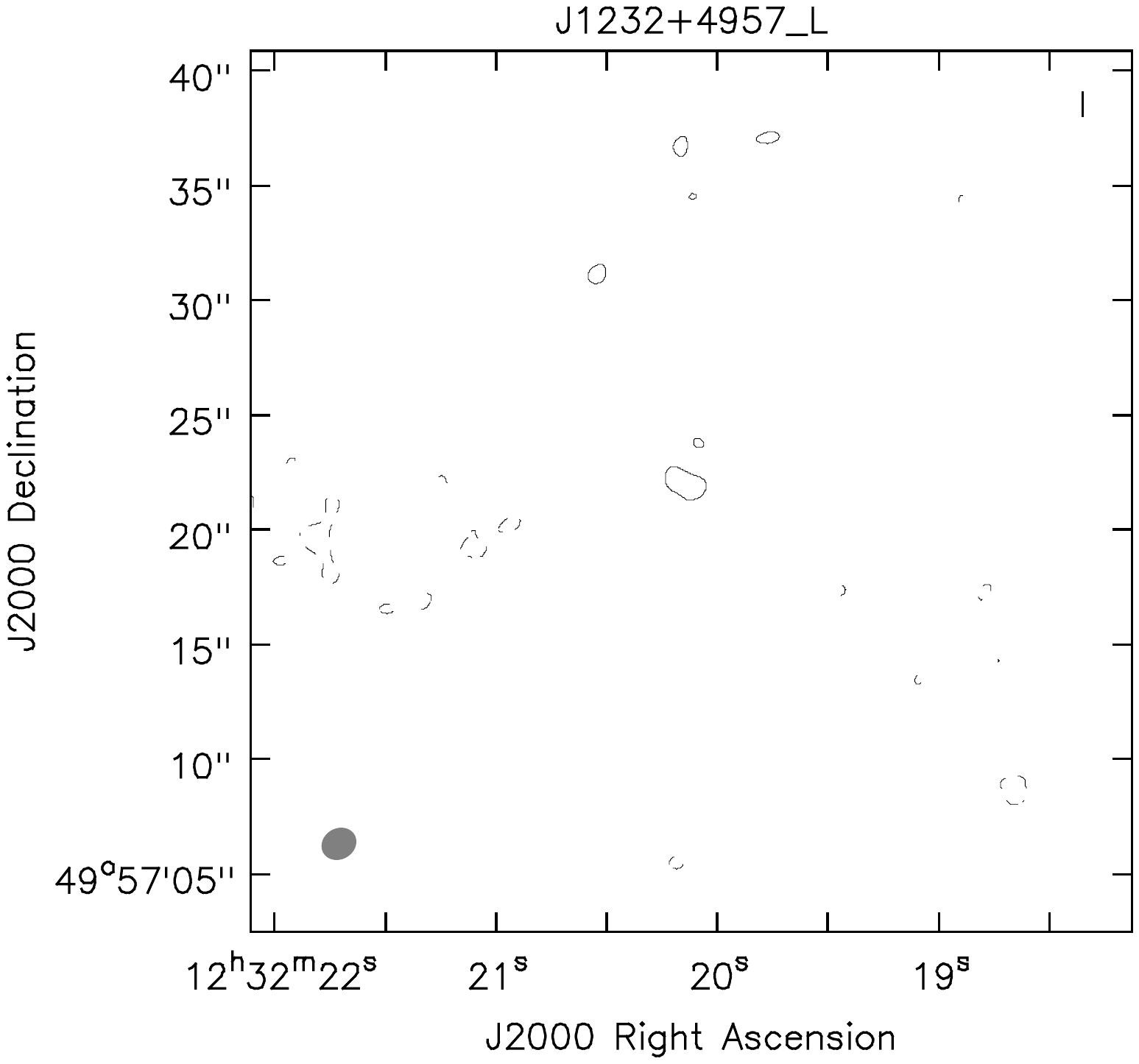}
    \end{minipage}
    \hfill
    \begin{minipage}[t]{.33\textwidth}
        \centering
	\includegraphics[trim={0cm 11cm 4cm 0cm}, width=\textwidth]{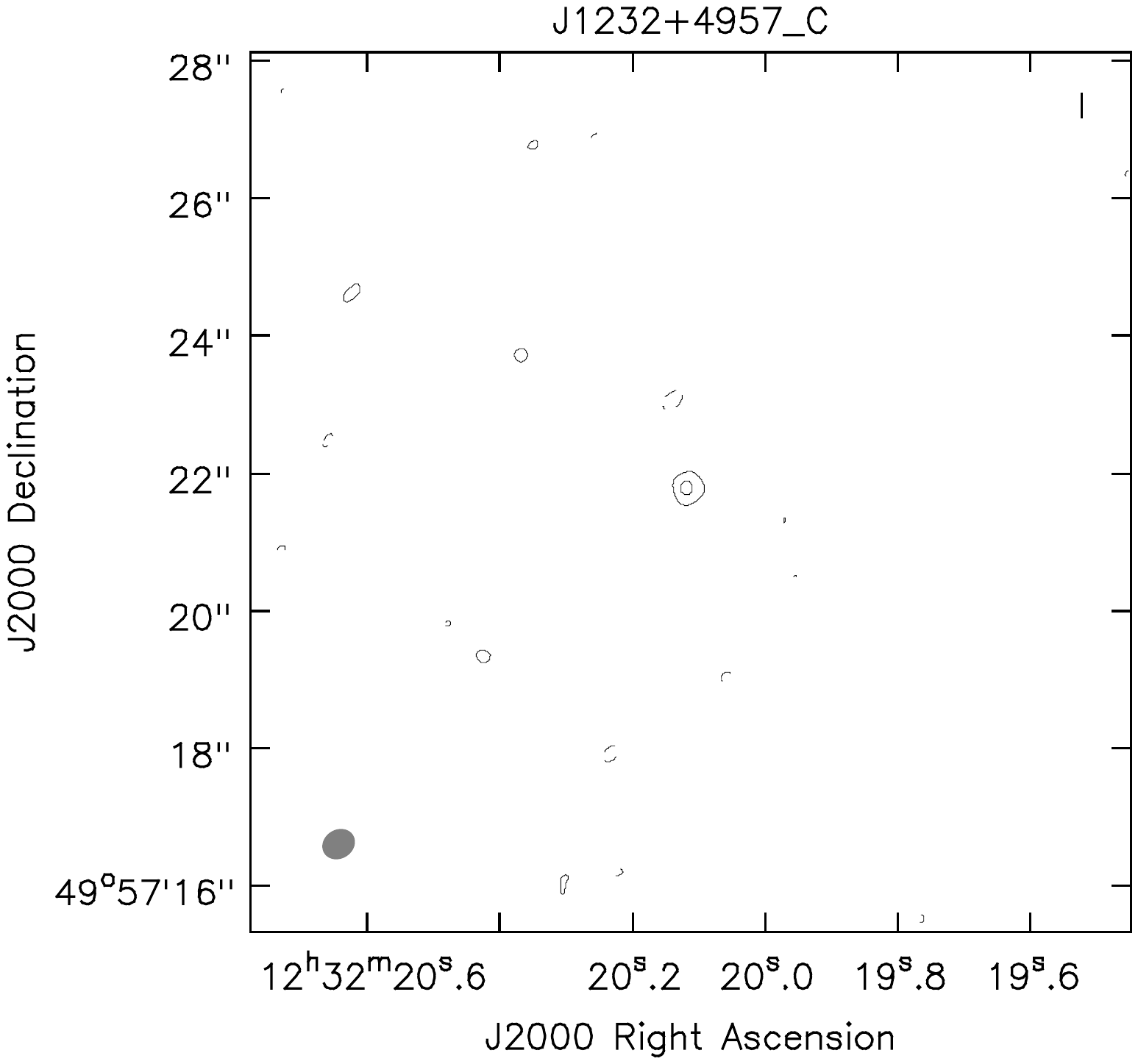}
    \end{minipage} 
    \hfill
    \begin{minipage}[t]{.33\textwidth}
        \centering
	\includegraphics[trim={0cm 11cm 4cm 0cm}, width=\textwidth]{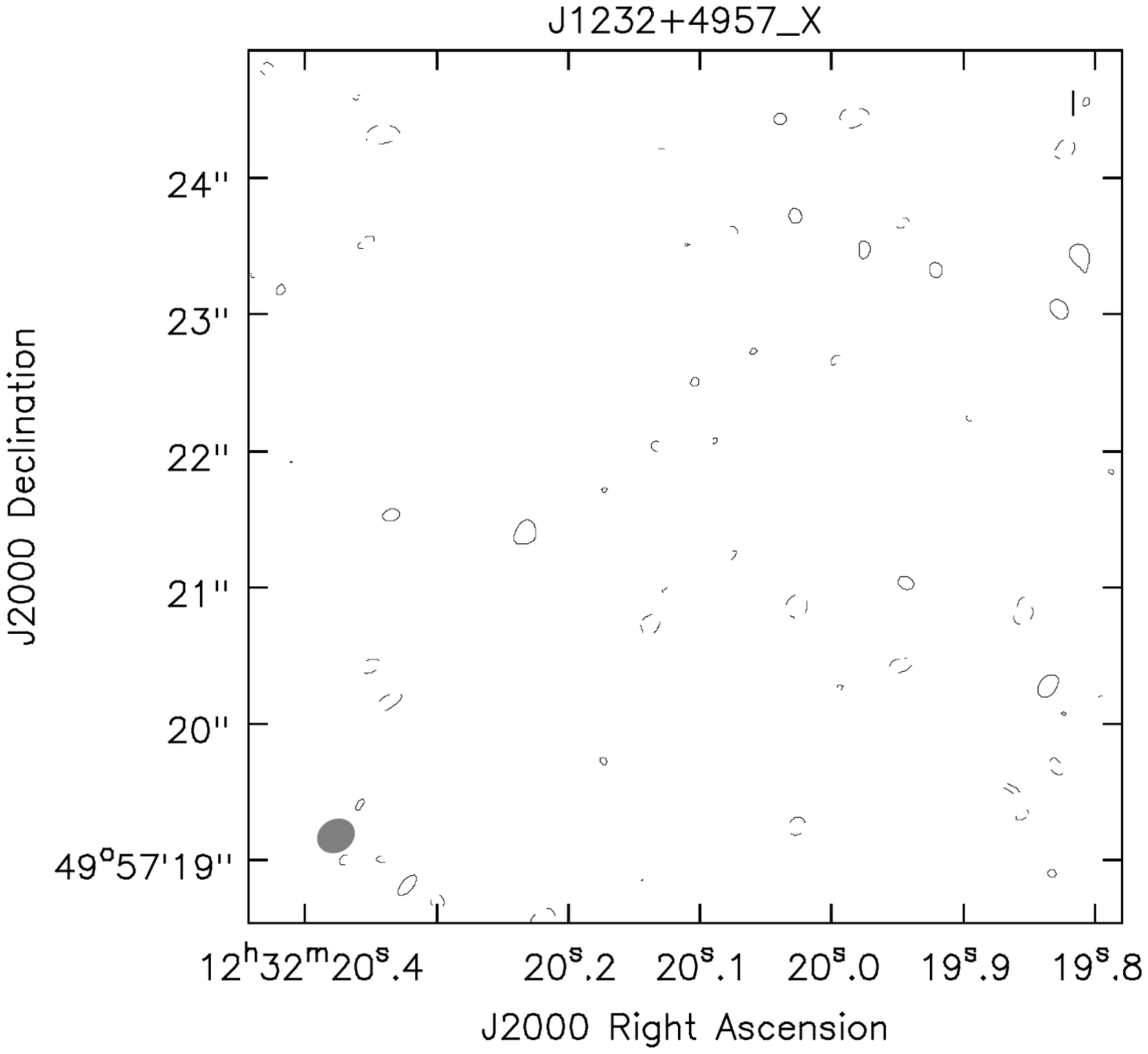}
    \end{minipage}  
\caption{From left to right, maps of J1232+4957 at 1.6 (L), 5.2 (C), and 9.0~GHz (X), respectively. The rms of the maps is 21, 7, and 7 $\mu$Jy, respectively, the contour levels are at rms$\times$(-3, 3$\times$2$^n$), n $\in$ [0,1], beam size 2.76$\times$2.25 kpc.}
\label{fig:J1232}
\end{figure*}

\begin{figure*}[t]
    \centering
    \begin{minipage}[t]{.33\textwidth}
	\centering
	\includegraphics[trim={0cm 11cm 4cm 0cm}, width=\textwidth]{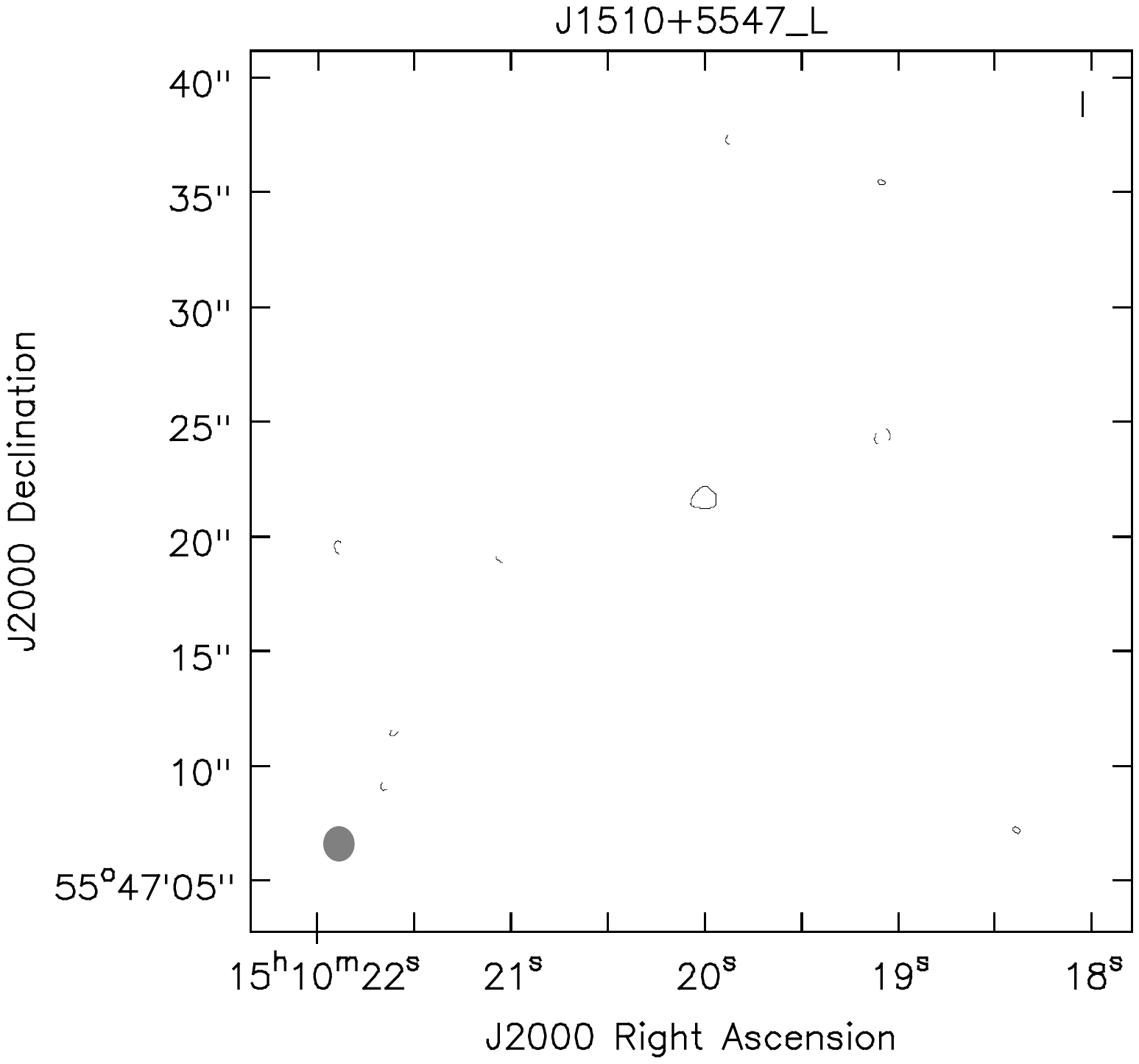}
    \end{minipage}
    \hfill
    \begin{minipage}[t]{.33\textwidth}
        \centering
	\includegraphics[trim={0cm 11cm 4cm 0cm}, width=\textwidth]{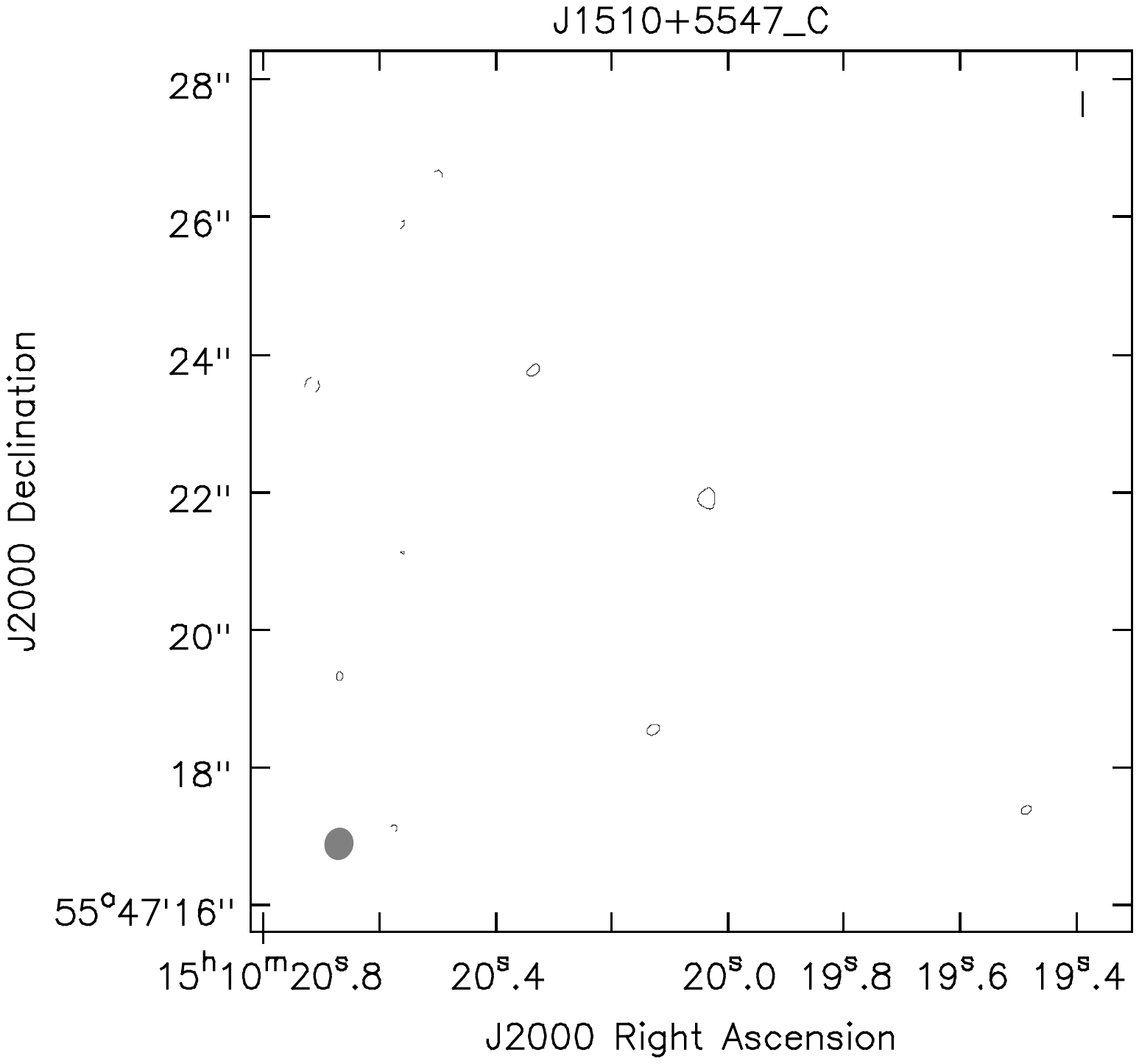}
    \end{minipage} 
    \hfill
    \begin{minipage}[t]{.33\textwidth}
        \centering
	\includegraphics[trim={0cm 11cm 4cm 0cm}, width=\textwidth]{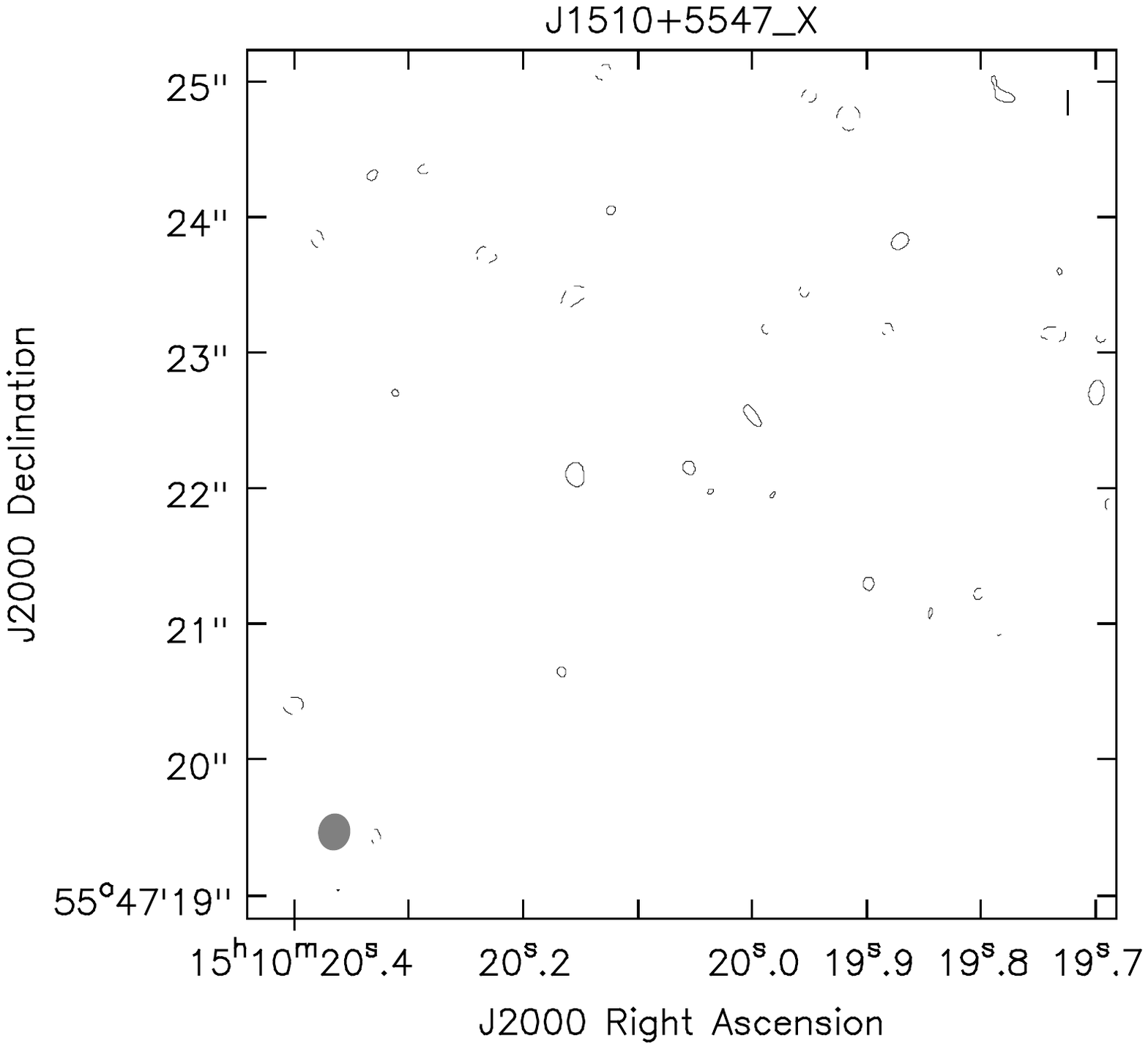}
    \end{minipage}  
\caption{From left to right, maps of J1510+5547 at 1.6 (L), 5.2 (C), and 9.0~GHz (X), respectively. The rms of the maps is 21, 8, and 7 $\mu$Jy, respectively, the contour levels are at rms$\times$(-3, 3), beam size 1.45$\times$1.29 kpc. }
\label{fig:J1510}
\end{figure*}

\begin{figure*}[t]
    \centering
    \begin{minipage}[t]{.33\textwidth}
	\centering
	\includegraphics[trim={0cm 11cm 4cm 0cm}, width=\textwidth]{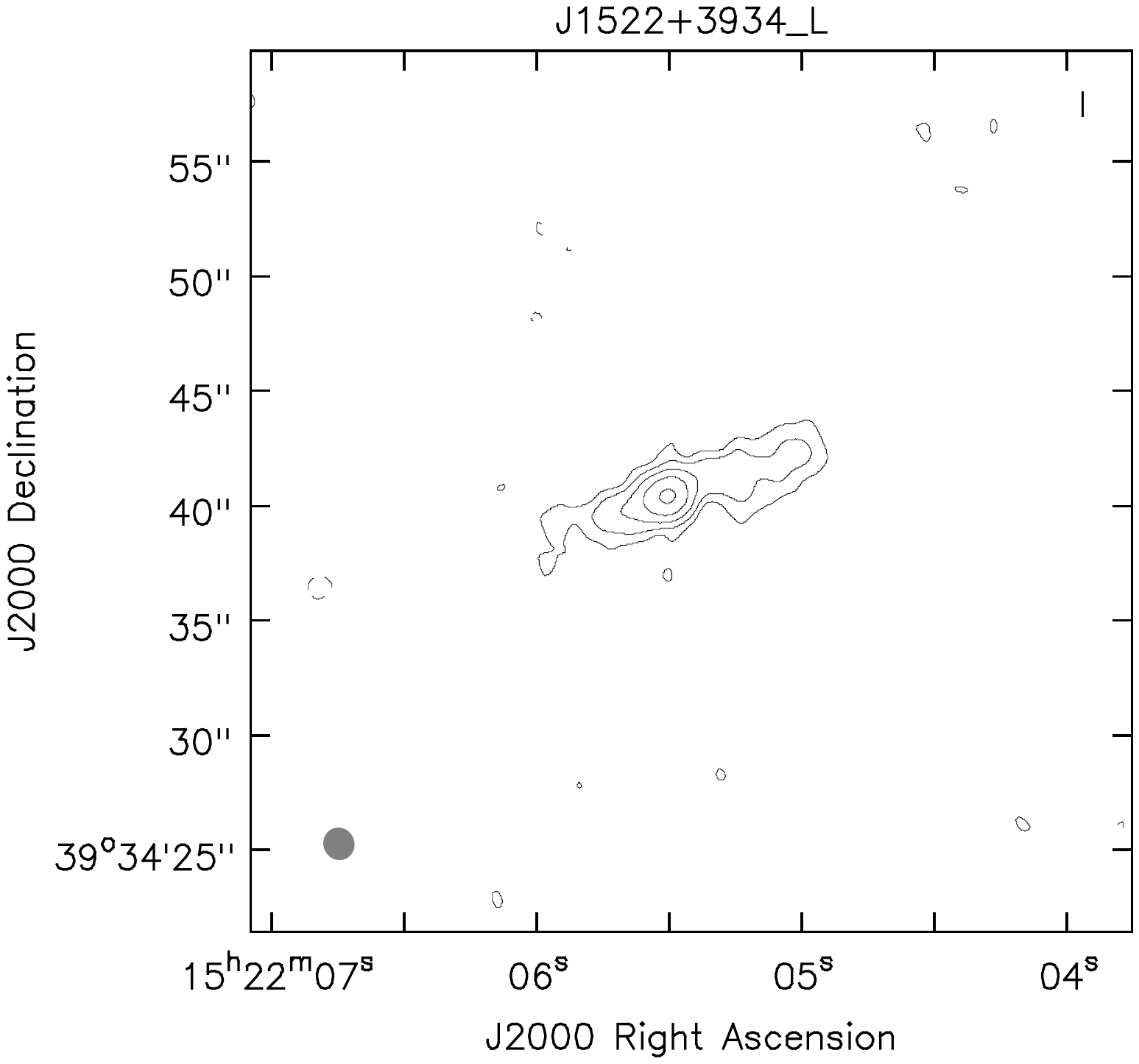}
    \end{minipage}
    \hfill
    \begin{minipage}[t]{.33\textwidth}
        \centering
	\includegraphics[trim={0cm 11cm 4cm 0cm}, width=\textwidth]{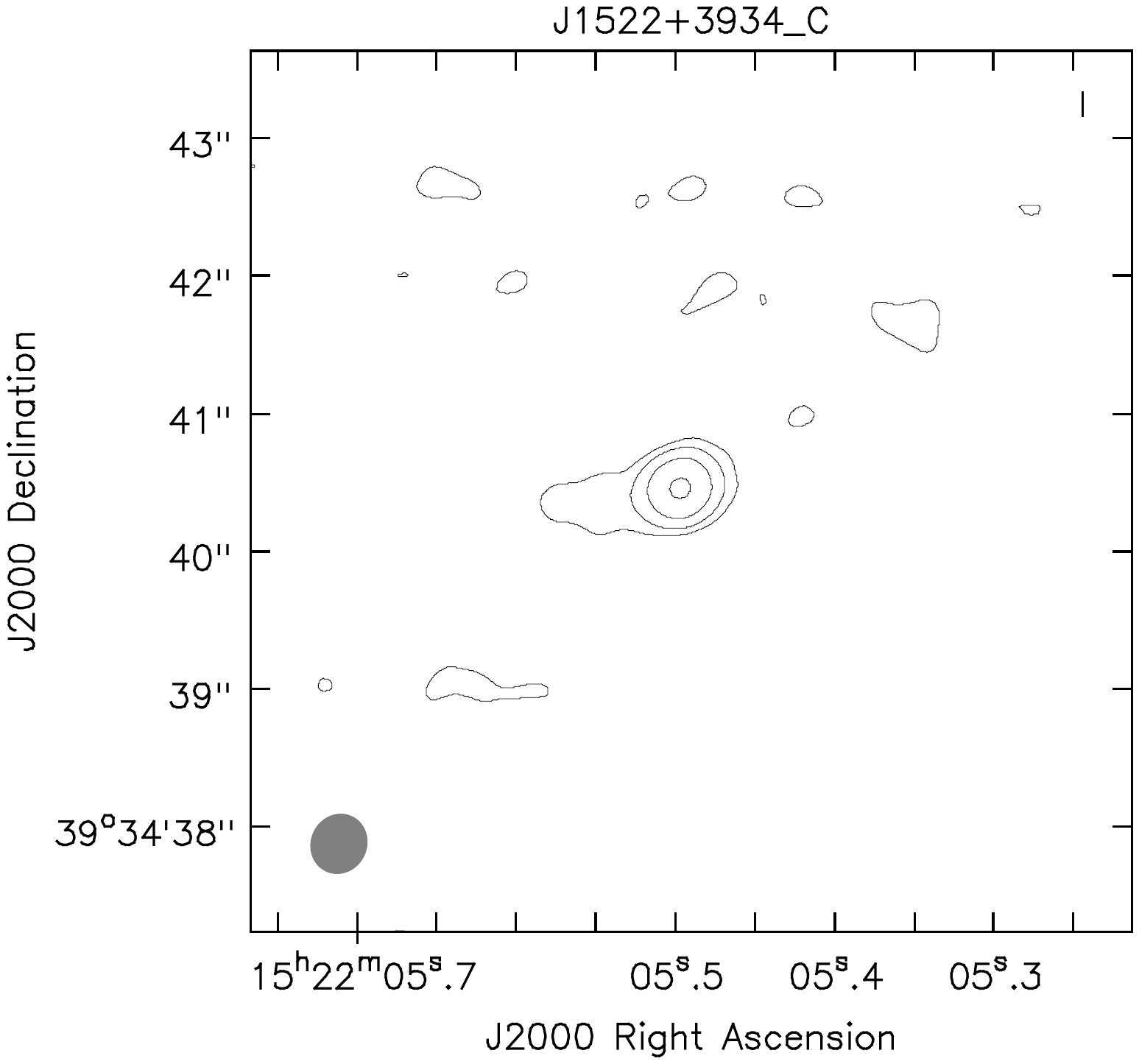}
    \end{minipage} 
    \hfill
    \begin{minipage}[t]{.33\textwidth}
        \centering
	\includegraphics[trim={0cm 11cm 4cm 0cm}, width=\textwidth]{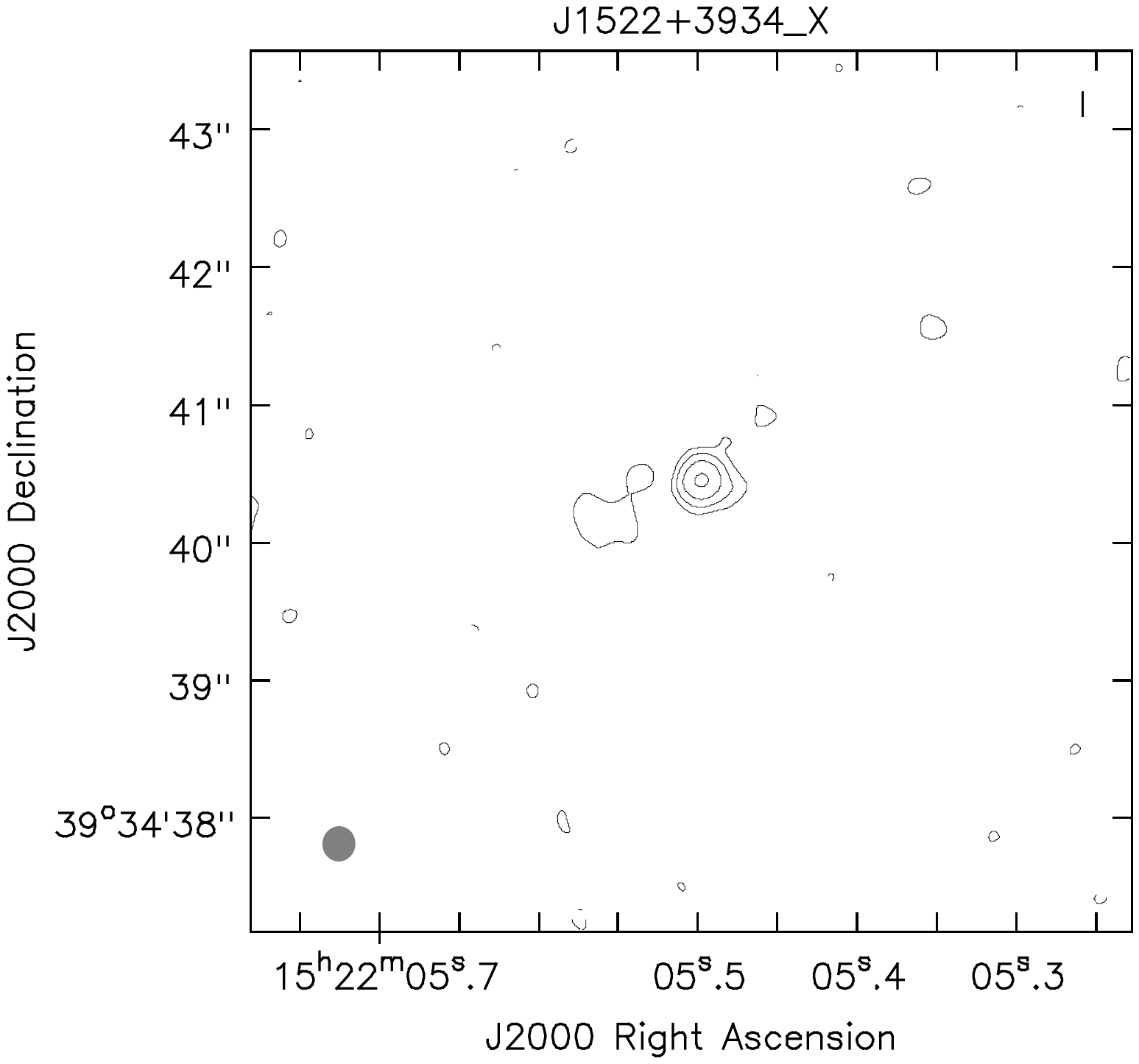}
    \end{minipage}  
\caption{From left to right, maps of J1522+3934 at 1.6 (L), 5.2 (C), and 9.0~GHz (X), respectively. The rms of the maps is 21, 12, and 7 $\mu$Jy, respectively, the contour levels are at rms$\times$(-3, 3$\times$2$^n$), n $\in$ [0,4], beam size 0.73$\times$0.66 kpc.}
\label{fig:J1522}
\end{figure*}

\begin{figure*}[t]
    \centering
    \begin{minipage}[t]{.33\textwidth}
	\centering
	\includegraphics[trim={0cm 11cm 4cm 0cm}, width=\textwidth]{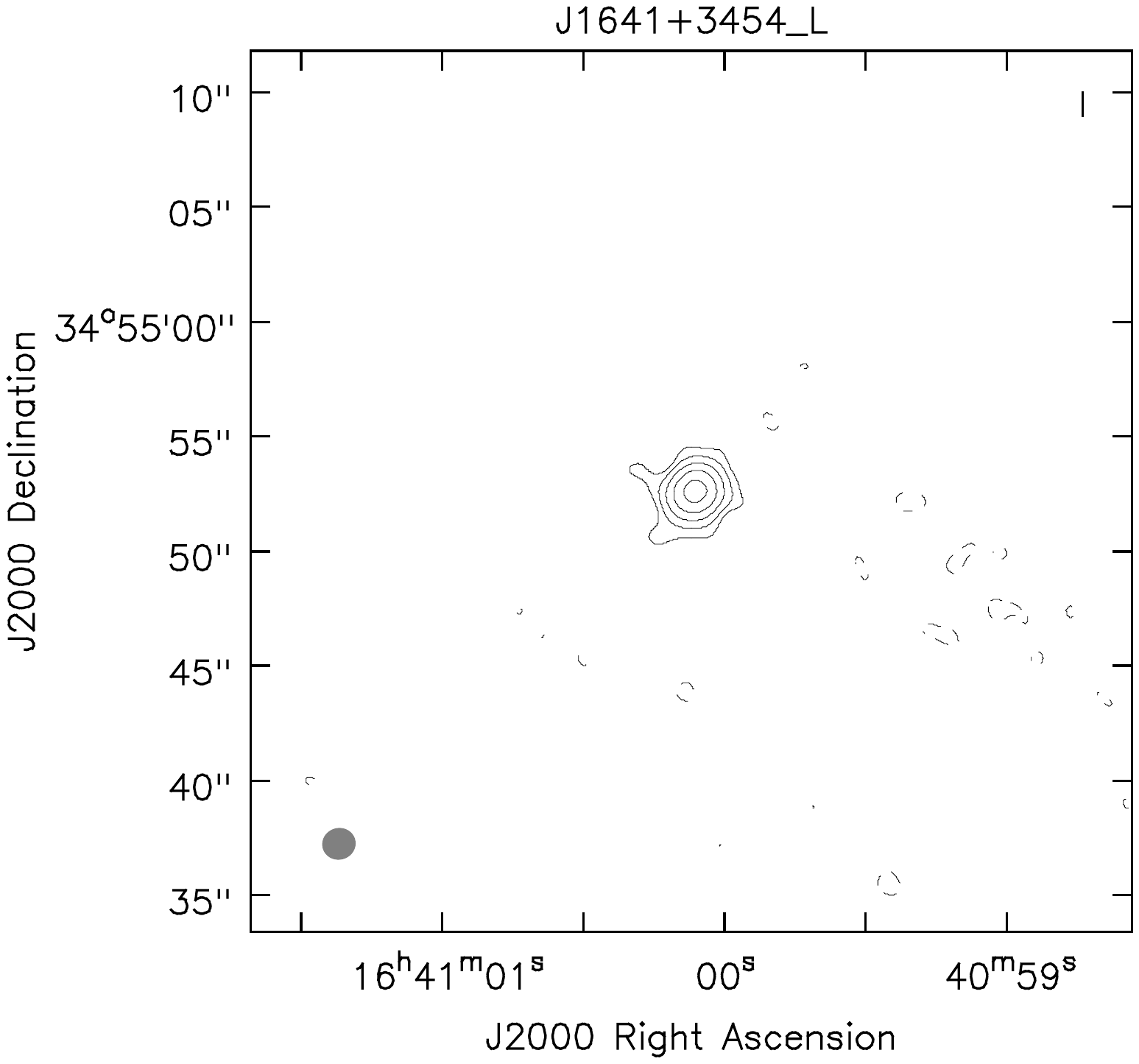}
    \end{minipage}
    \hfill
    \begin{minipage}[t]{.33\textwidth}
        \centering
	\includegraphics[trim={0cm 11cm 4cm 0cm}, width=\textwidth]{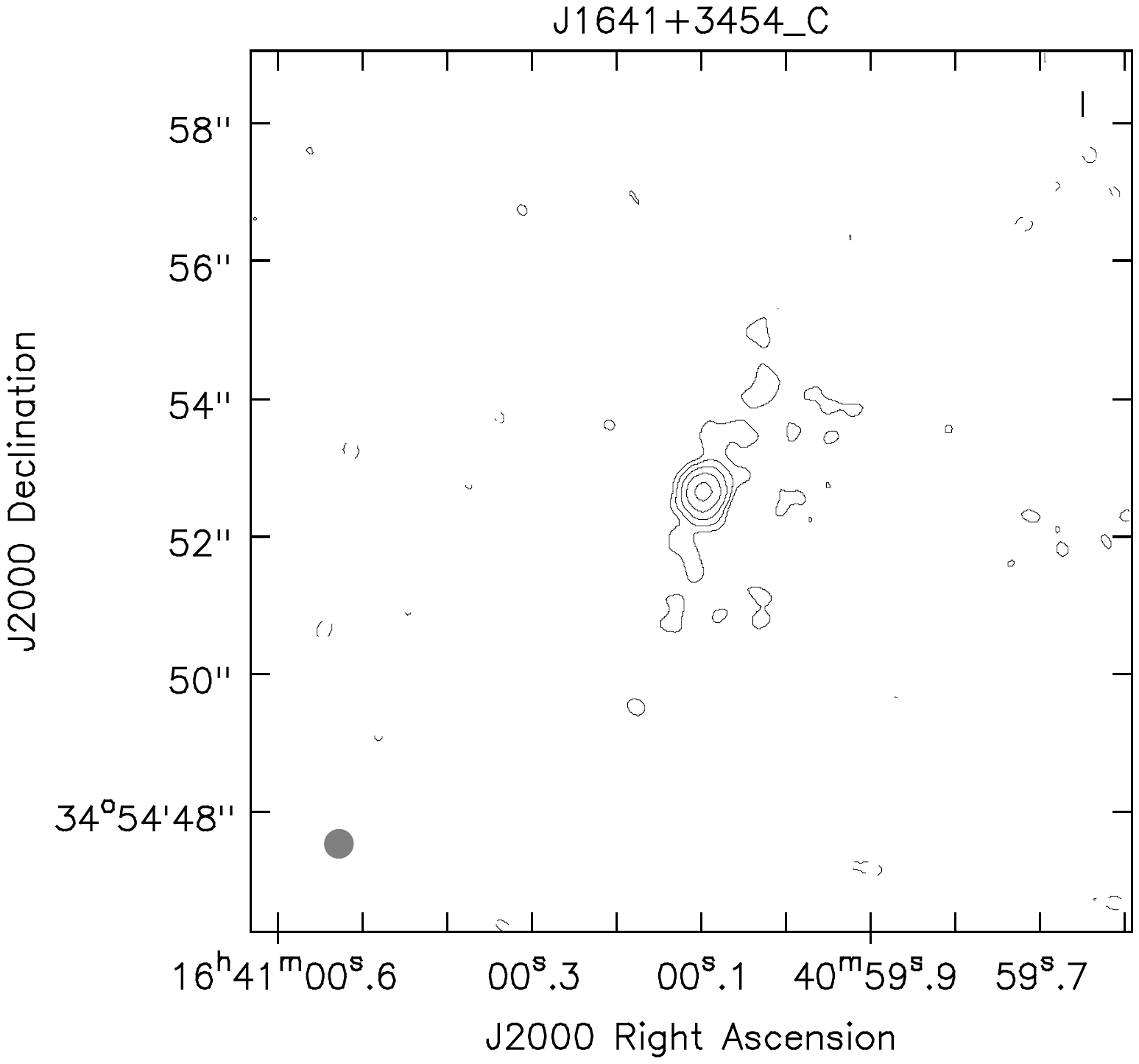}
    \end{minipage} 
    \hfill
    \begin{minipage}[t]{.33\textwidth}
        \centering
	\includegraphics[trim={0cm 11cm 4cm 0cm}, width=\textwidth]{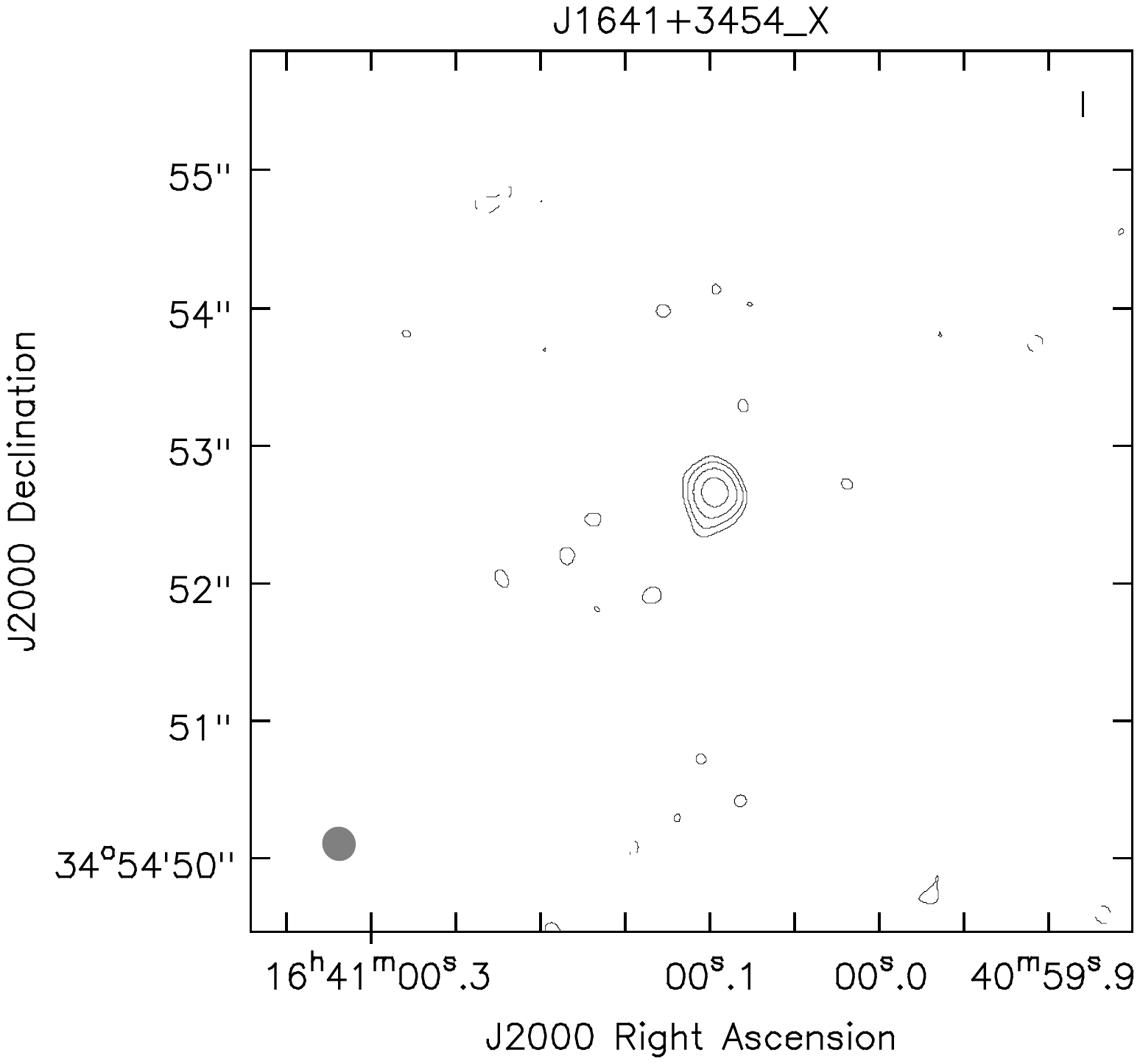}
    \end{minipage}  
\caption{From left to right, maps of J1641+3454 :joy:. The rms of the maps is 27, 8, and 7 $\mu$Jy, respectively, the contour levels are at rms$\times$(-3, 3$\times$2$^n$), n $\in$ [0,4], beam size 1.48$\times$1.44 kpc.}
\label{fig:J1641}
\end{figure*}

To produce maps, when a source was detected we modeled it using a single Gaussian in the image plane and we deconvolved it from the core, hence deriving the core deconvolved size, its position angle (PA), the peak and integrated flux density.
If the source was not point-like, we measured the integrated flux density in the region within the 3$\sigma$ contour, where $\sigma$ is the noise of the image. 
We then calculated the luminosity accounting for K-correction by following
\begin{equation}
L = 4\pi d_l^2 \nu S_\nu (1 + z)^{\alpha_\nu - 1} \; ,
\end{equation}
where $S_\nu$ is the flux density at frequency $\nu$, and $d_l$ is the luminosity distance corresponding to redshift $z$, and for $\alpha_\nu$ we used the spectral indexes calculated from our data and shown in Table \ref{tab:measurements}.
The spectral indexes between 1.6~GHz and 5.2~GHz and between 5.2~GHz and 9.0~GHz were estimated following the usual relation
\begin{equation}
\alpha_{1,2} = -\frac{\log\left(\frac{S_1}{S_2}\right)}{\log\left(\frac{\nu_1}{\nu_2}\right)} \; .
\end{equation}
Finally, we estimated the brightness temperature of the sources as done by \citet{Doi13},
\begin{equation}
T_b = 1.8\times 10^9 (1 + z) \frac{S_{\nu,p}}{\nu^2 \theta_{maj} \theta_{min}} [\mathrm{K}] \; ,
\end{equation}
where $S_{\nu,p}$ is the peak flux density expressed in mJy beam$^{-1}$, $\nu$ is the frequency in GHz, and $\theta_{maj}$, $\theta_{min}$ are the major and minor axis of the source core in milliarcsec, deconvolved from beam. 
It is worth noting that usually the brightness temperature is estimated from high-resolution measurements, therefore our estimates will only be lower limits. 
However, they will be directly comparable with those already derived for different classes of NLS1s by \citet{Berton18a}, whose observations were also carried out with the JVLA in A configuration.
For this reason, we limit our brightness temperature estimate at 5.2~GHz, where the sources in \citet{Berton18a} were studied. 
The radio maps of the detected sources are shown in Figures \ref{fig:J1228} to \ref{fig:J1641}. 
Flux densities and luminosities are reported in Table \ref{tab:fluxes}, while other physical parameters we measured from the maps are shown in Table \ref{tab:measurements}.

\section{Results}

Out of the seven sources we observed, four were detected in all three bands with JVLA. 
One of them, J1232+4957, was detected only at 1.6~GHz and 5.2~GHz. 
The remaining two, J1029+5556 and J1509+6137, were not detected in any band. \par
Two of the sources, J1522+3934 and J1641+3454, have been already observed in the FIRST survey at 1.6~GHz in B configuration, which measured flux densities of 1.88$\pm$0.14 and 2.43$\pm$0.13 mJy, respectively. 
The integrated flux density we measured at 1.6~GHz was 2.03$\pm$0.16 and 2.52$\pm$0.10 mJy, respectively, results well in agreement with those from the FIRST.
Regarding the other sources, all of them are still below the detection threshold of the FIRST. 
In conclusion, we do not find any significant luminosity increase in the last $\sim$25 years for any of our sources at 1.6~GHz. \par
\begin{figure}[t!]
\includegraphics[width=\hsize]{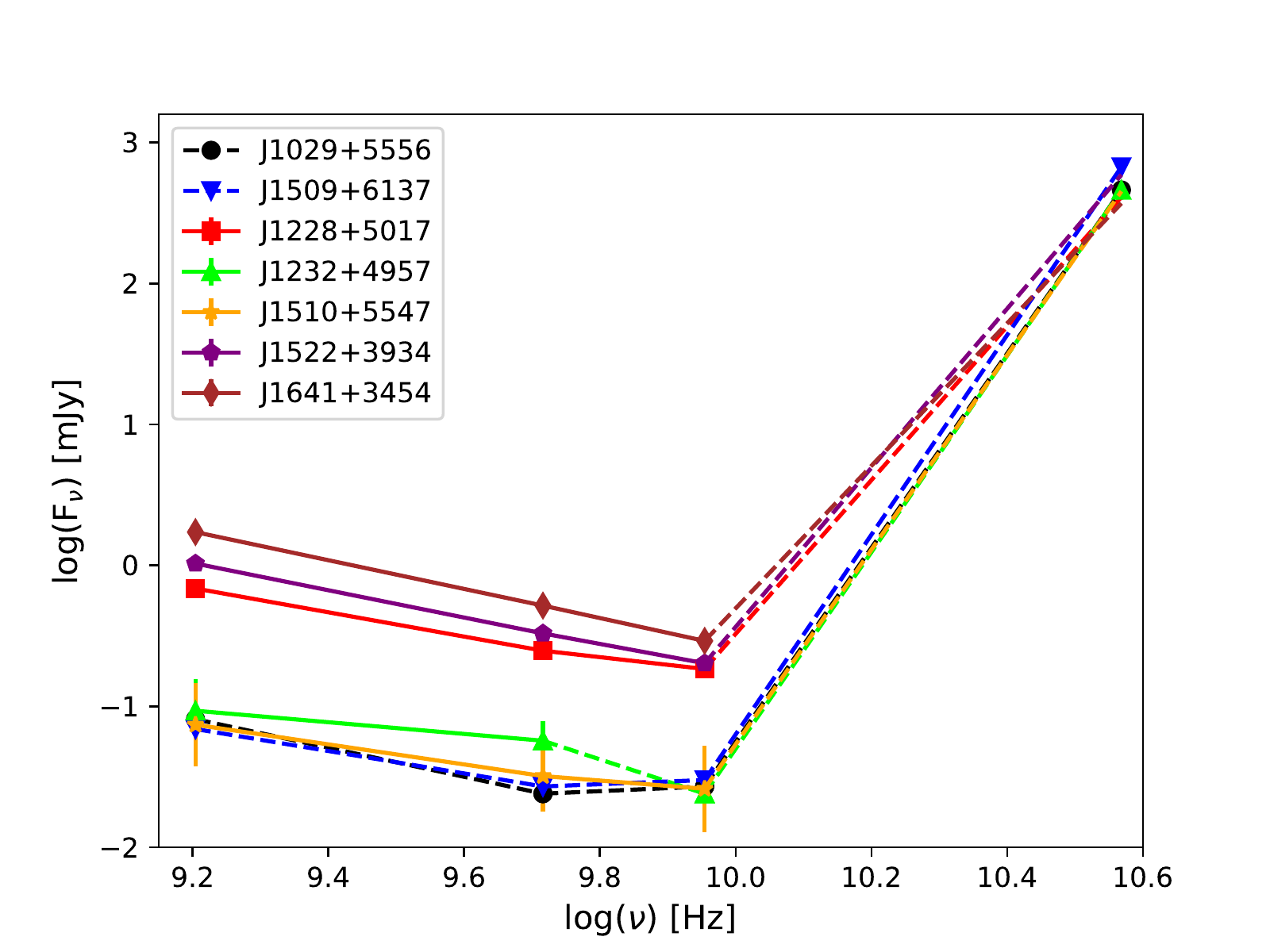}
\caption{The radio spectra of our sample. The three points below 10 GHz are those measured by the JVLA, while the last one is the average value detected at 37~GHz. We remark that the last point was not observed simultaneously to the others. The solid lines represent the spectrum, the dashed lines are connecting either upper or lower limits of flux densities, therefore they represent limits on the spectral index. }
\label{fig:spectra}
\end{figure}

Our observations at 5.2~GHz and 9.0~GHz are the first ever performed for these objects. 
The observed spectra of all the sources is shown in Fig.~\ref{fig:spectra}. 
The spectral indexes we measured clearly indicate that the spectrum remains steep in the whole spectral region covered by our JVLA data.
After that, at some point above 10~GHz, the spectra seem to change slope and start rising, to reach the flux densities measured at 37~GHz. 
Between 1.6~GHz and 5.2~GHz, the median spectral index of the peak flux is 0.86, while between 5.2~GHz and 9.0~GHz the median is 0.71. 
Therefore, the slope tends to become flatter already below 10 GHz. 
This behavior is observed in three out of five sources with known spectral index. 
Only in J1641+3454 the spectral indexes remain the same, while J1232+4957 is not detected at 9.0~GHz and the spectral index is only an upper limit. 
In terms of integrated emission, the behavior stays the same.
The median spectral index of the total emission is 1.31 between 1.6~GHz and 5.2~GHz, and 0.82 between 5.2~GHz and 9.0~GHz. 
As before, J1614+3454 has a constant spectral index, while J1232+4957 becomes steeper. 
When our results are compared to the in-band spectral indexes measured in at 5.2~GHz by \citet{Berton18a}, they turn out to be comparable either to those of typical radio-quiet and putatively non-jetted NLS1s, or with those of jetted steep-spectrum radio-loud NLS1s \citep[i.e. misaligned objects, see][]{Berton16c}.
However, only J1641+3454 could be formally classified as radio-loud, while all the other objects in our sample are radio-quiet \citep{Lahteenmaki18}. 
We did not calculate the spectral index of our sources between 9.0~GHz and 37~GHz because the data are not simultaneous, and the 37~GHz flux is likely measured during flares. 
It is worth noting that above 10 GHz the spectra are somehow similar to those observed in a handful of $\gamma$-ray emitting NLS1s, such as 1H 0323+342 and FBQS J1644+2619 \citep{Lahteenmaki17}. 
In analogy to what is seen here, also in those cases the spectrum tends to rise at higher frequencies, although with significantly less extreme spectral indexes.\par
As mentioned above, we estimated the brightness temperature of our sources using only data at 5.2~GHz. 
The median value is approximately $\sim$100 K, while the highest temperature limit (831 K) is found in J1641+3454, which as mentioned above is the only object of our sample both radio-loud and detected in $\gamma$ rays. 
Nevertheless, when compared to the samples by \citet{Berton18a}, the brightness temperature limit of J1641+3454 is lower than that of any beamed object, but it is within the range of both radio-quiet and steep-spectrum radio-loud sources. 
Regarding the remaining four sources with a measured brightness temperature, two of them lie below the threshold of 100 K, where only radio-quiet sources are found, while the other two, albeit having a lower temperature than J1641+3454, are also found in the region where both radio-quiet and steep-spectrum radio-loud objects are. \par
Finally, in terms of luminosity, \citet{Berton18a} found no jetted source with integrated luminosity below $1.5\times10^{39}$ \ergs\ at 5.2~GHz. 
The brightest of our sources is J1228+5017, which has a luminosity at 5.2~GHz of $1.0\times10^{39}$ \ergs\, already below that threshold. 
All the other objects, of course, have even lower luminosities, including the sources where only upper limits could be estimated.
Therefore, all the NLS1s in our sample seem to be more related to typical radio-quiet sources than to steep-spectrum radio-loud objects.

%\begin{figure*}[ht]
%    \centering
 %   \begin{minipage}[t]{.33\textwidth}
%	\centering
%	\includegraphics[trim={0cm 11cm 4cm 0cm}, width=\textwidth]{./data/J1029/J1029_L.pdf}
%    \end{minipage}
%    \hfill
%    \begin{minipage}[t]{.33\textwidth}
%        \centering
%	\includegraphics[trim={0cm 11cm 4cm 0cm}, width=\textwidth]{./data/J1029/J1029_C.pdf}
%    \end{minipage} 
%    \hfill
%    \begin{minipage}[t]{.33\textwidth}
%        \centering
%	\includegraphics[trim={0cm 11cm 4cm 0cm}, width=\textwidth]{./data/J1029/J1029_X.pdf}
%    \end{minipage}  
%\caption{From left to right, maps of J1029+5556 in L, C, and X band, respectively. }
%\label{fig:J1029}
%\end{figure*}

%\begin{figure*}[t]
%    \centering
%    \begin{minipage}[t]{.33\textwidth}
%	\centering
%	\includegraphics[trim={0cm 11cm 4cm 0cm}, width=\textwidth]{./data/J1509/J1509_L.pdf}
%    \end{minipage}
%    \hfill
%    \begin{minipage}[t]{.33\textwidth}
%        \centering
%	\includegraphics[trim={0cm 11cm 4cm 0cm}, width=\textwidth]{./data/J1509/J1509_C.pdf}
%    \end{minipage} 
%    \hfill
 %   \begin{minipage}[t]{.33\textwidth}
  %      \centering
%	\includegraphics[trim={0cm 11cm 4cm 0cm}, width=\textwidth]{./data/J1509/J1509_X.pdf}
 %   \end{minipage}  
%\caption{From left to right, maps of J1509+6137 in L, C, and X band, respectively. }
%\label{fig:J1509}
%\end{figure*}

\section{Discussion}

\subsection{A convex radio spectrum in NLS1s}
\label{sec:SSA}
Considering the results of our JVLA observations, the detection of radio-quiet/silent NLS1s at 37~GHz becomes even more puzzling.
If the spectra would keep decreasing with the spectral index we measured between 5.2 and 9.0~GHz, in the best-case scenario (that of J1641+3454) the expected flux density at 37~GHz would be 82 $\mu$Jy. 
Instead, we are observing a flux density $\sim$4500 times higher than that. 
Even for the flare of a jetted AGN, such luminosity increase would be fairly extreme, and it would reflect also at other wavelengths such as in optical.
However, none of our sources have been detected in a flaring state at optical wavelengths by any survey aimed at finding transient objects. 
Therefore, it is not far-fetched to assume that the spectrum is not decreasing linearly, but that it is instead convex. 
The position of the minimum cannot be determined from our data. 
However, we can hypothesize that it is located at some point between 9.0 and 37~GHz. 
The spectral indexes between these bands, using the data measured at 37~GHz, are between $\alpha_{9-37 \; GHz} = -5.32$ for J1641+3454 and $\alpha_{9-37 \; GHz} = -6.8$ in the case of J1510+5547. 
Such values are not measured with simultaneous observations, therefore they are almost certainly lower limits. 
It is likely that the 37~GHz observations were carried out during a flaring state of the source.
Indeed, the detection of these sources at 37~GHz clearly indicates that the high-frequency radio emission of these NLS1s is highly variable. 
The fluxes measured in different epochs at 37~GHz are not consistent with each other \citep{Lahteenmaki18}, and the most likely explanation for such a strong variability (J1522+3934 went from 0.3 Jy to 1.4 Jy within a few weeks) is the presence of a relativistic jet (although an alternative option will be explored in the following, see Sect.~\ref{sec:corona}). \par
An inverted spectrum from a relativistic jet may originate due to synchrotron self-absorption. 
Assuming its typical spectral index, $\alpha = -2.5$, and extrapolating from the flux density measured at 37~GHz, the expected $S_\nu$ at 9.0~GHz during the flare should be around 13 mJy.
This value corresponds to a flux density increase of $\sim 400$ times which, under typical conditions, should also be seen by optical transient surveys \citep[e.g.,][]{Bellm19}, assuming that the flare occurs throughout the whole electromagnetic spectrum. 
Therefore, the spectrum between 9 and 37~GHz may have a higher slope than that of synchrotron self-absorption, and the high-frequency excess is not (entirely) due to a flaring state of the source.\par

\subsection{Star formation at low frequencies}

If a relativistic jet is present, it should be visible also at low frequencies. 
However, JVLA observations do not seem to confirm the presence of this feature. 
The radio spectrum of most of our sources, indeed, can be perfectly explained without any need for a jet component. 
All of them, in terms of spectral index, brightness temperature, and especially luminosity, seem to be dominated by star formation. 
The latter is a combination of free-free emission, which has an almost flat radio spectrum, and steep synchrotron emission from supernova remnants. 
When combined, these lead to an observed spectral index of $\sim$0.7, which is characteristic of star-forming galaxies, and roughly in agreement with the spectral indexes of our sources \citep[e.g.,][]{Condon92, Panessa19}. \par

However, at least in the $\gamma$-ray source J1641+3454, a relativistic jet must be present.
The detection of $\gamma$ rays is associated with a beamed relativistic jet which, typically, has a flat radio spectrum in the JVLA range. 
However, its low-frequency spectrum is clearly steep, its brightness temperature is rather low (albeit still consistent with a misaligned jet), and its radio luminosity is very low, the lowest ever measured for a $\gamma$-ray NLS1 \citep[e.g.,][]{Berton18a, Paliya19a}. 
Even in this object, therefore, the radio emission in this spectral region seems not to be dominated by the relativistic jet, but from star formation activity. \par
In population A sources of the quasar main sequence, especially those with prominent Fe II features, the star formation is typically very strong \citep{Sani10, Marziani18b}. 
In some cases, its contribution to the radio emission can be so important that the source is classified as radio-loud even if the relativistic jet is not present \citep{Ganci19}. 
Objects like this may be common among NLS1s. 
The infrared colors of some flat-spectrum radio-loud sources seem to suggest that the bulk of their radio emission originates from star forming regions \citep{Caccianiga15}. 
Therefore, it would not be surprising to find out that even the radio emission of a (barely) radio-loud source such as J1641+3454 is due to star formation.
It is also worth noting that the 1.6~GHz map of J1522+3934 of Fig.~\ref{fig:J1522} (left) has a nice counterpart in the infrared morphology of the source \citep{Jarvela18}. 
This source, indeed, is hosted by a spiral galaxy which is interacting with a close companion, and the diffuse radio emission is well aligned with the two nuclei, possibly because it originates in a starburst induced by the ongoing interaction. \par
However, we need to consider whether star formation could account for the $\gamma$-ray emission of J1641+3454, and if the prominent variability observed at high frequency at 37~GHz could be explained if only star formation was present. \par % NO.
We estimated the radio flux densities produced by star formation processes by using the known relation between the radio and infrared emission. We utilized the equations given in \citet{Boyle07} for the 1.5~Ghz -- 24~$\mu$m relation. Instead of 24~$\mu$m data we used the Wide-field Infrared Survey Explorer (WISE) W4 data that has a wavelength of 22~$\mu$m, and our radio observations are at 1.6~GHz, but these small differences should not significantly affect the relation. The radio flux densities estimated from the WISE W4 flux densities are well in agreement with the values observed at 1.6~GHz, except for two sources, J1522+3934 and J1641+3454. The radio flux density of J1522+3934 is almost an order of magnitude higher than what the W4 flux density predicts, and additionally its W4 flux density value has a variability flag, suggesting the jet contribution might not be insignificant at 22~$\mu$m. However, despite the possible contribution of the jet its radio flux density is higher than expected, indicating that star formation is not a sufficient explanation for the whole radio emission. J1641+3454 is not that extreme as the observed radio emission is about double compared to the predicted one. There can be a jet contribution at 1.6~GHz, increasing toward higher frequencies, as indicated by the slightly flattening spectral indices between 5.2 and 9.0~GHz when compared to the one between 1.6 and 5.2~GHz.

\subsection{A coronal emission scenario}
\label{sec:corona}
The presence of an inverted slope toward high frequency in the radio spectrum of AGN is not entirely new. 
Already in the eighties some observations of radio-quiet quasars found out that in some sources an excess at high frequencies is present, particularly in type 1 AGN \citep{Edelson87, Antonucci88, Barvainis89}. 
Similar high frequency excesses have been often found in the literature among radio-quiet quasars \citep{Planck11, Behar15, Doi16a}, although none of them have the same extremely variable behavior as our sources. \par
An interesting possibility instead is that the radio emission originates from the AGN corona, which as seen from X-rays is rather prominent in NLS1s \citep{Gallo18}. 
Since the corona is very compact, \citet{Laor08} suggested that its emission would be synchrotron self-absorbed in the low-frequency range where star formation emission dominates instead, and therefore it could appear as a sudden change of slope in the spectrum, with a peak around some hundreds of GHz. 
In particular, due to its nature, this emission would be strongly variable, and correlate with the X-ray emission. \par
Recently, we found that shortly after a new radio detection of J1641+3454 at 37~GHz the source showed a brightening in the X-rays with respect to its luminosity measured in the low state.
A detailed analysis of the properties of J1641+3454 in X-rays will be the subject of an upcoming paper.
However, this result already confirms that the radio emission detected at 37~GHz is actually coming from J1641+3454. 
Such behavior could be consistent with what is expected if the corona is responsible from the radio emission. 
However, it is not different from what is observed during regular blazar flares, in which both the radio and the X-rays originate in a relativistic jet \citep[e.g.,][]{Berton18b}. 
Therefore, this observation does not prove the coronal origin of the high-frequency emission. 
Furthermore, the luminosity we measure at 37~GHz is at least one order of magnitude larger than any previous measurement from radio-quiet quasars. \par
To test this hypothesis, we checked the position of our sources on the G\"udel-Benz (GB) relation for radio-quiet quasars \citep{Gudel93}, that links the X-ray emission to the radio emission. In principle, we would expect a roughly constant ratio between the radio and the X-ray luminosities, with $L_R \sim 10^{-5} L_x$. We derived the X-ray luminosity for 5 out of 7 sources from the ROSAT archive \citep{Boller16}. Two sources, J1029+5556 and J1522+3934, have no X-ray detections. Interestingly, when tested against the luminosities at 1.6~GHz, our sources lie close to the relation. However, when the X-ray luminosity is compared to the 37~GHz luminosity, all of our sources deviate significantly from the GB relation, with ratios between 0.1 and 1. This seems to indicate that the 37~GHz emissions do not share the same origin, and that therefore the corona likely is not responsible for the high-frequency excess.
In conclusion, albeit appealing, the coronal origin of this radio emission may not be the correct explanation for this phenomenon. 

\subsection{Extremely young GPS and synchrotron self absorption}
\label{sec:GPS}
The classes of radio sources known as Gigahertz Peaked Sources (GPS) and high-frequency peakers (HFP) are kinematically young AGN observed at large angles.
It is widely believed that, with time, they will first grow into the other class of young sources known as compact steep-spectrum (CSS) sources, and then into fully developed radio galaxies \citep{Fanti95, Odea98}. 
Young radio sources are characterized by a spectral turnover which peaks around 1~GHz for GPS, at higher frequencies for HFP, or at lower frequencies for CSS sources.
The rising part of the spectrum is likely due to synchrotron self absorption (although free-free absorption may also play a role), while above the turnover the gas is optically thin and the synchrotron slope reflects the energy distribution of the electrons. 
The turnover frequency decreases when the projected linear size of the radio source increases, that is big sources peak at lower frequencies than small sources.
This behavior is typically explained in terms of kinematic age of the radio source.
Older sources are bigger than the young ones, their relativistic jets are expanding, their energy density decreases, and their spectra peak at lower and lower frequencies. 
Therefore, extremely young objects would peak at very high frequencies, as observed in HFP. \par
It is widely accepted that NLS1s, both jetted and non-jetted, are also AGN in an early phase of their evolution \citep{Mathur00, Komossa18, Paliya19a}. 
For this reason, some authors suggested that a link between NLS1s and CSS/GPS sources could exist \citep[e.g.][]{Oshlack01, Gallo06, Komossa06, Yuan08, Caccianiga14, Schulz15, Gu15, Caccianiga17, Liao20}, and even that CSS/GPS could constitute part of the parent population of jetted NLS1s \citep{Berton17, Foschini17}. 
If this is the case, some NLS1s may have only recently developed relativistic jets, which would be still confined within a few parsecs from the central engine. 
Using the relation by \citet{Odea97}
\begin{equation}
    \log\nu_{to} = -(0.21\pm0.05) - (0.65\pm0.05) \log LS \; ,
\end{equation}
where $\nu_{\rm to}$ is the turn-off frequency in GHz and $LS$ is the linear size in kpc, a source peaked at 37~GHz would correspond to a linear size of $\sim$1.8~pc. 
Assuming that the relativistic jet is propagating at the speed of light, this would in turn correspond to a kinematic age of the relativistic jet of 6~years. 
With a more realistic propagation velocity of $\sim 0.3$c, the age would be approximately 20 years, which would be low but not unprecedented \citep{Giroletti09a}.
It is also worth noting that a dense circumnuclear medium could also hamper the propagation of relativistic jets due its interaction with the clouds \citep{Vanbreugel84}.
This effect would be enhanced in presence of a non flattened broad-line region (BLR) where clouds are still located above and below the accretion disk plane, a geometry which may be realistic in the case of NLS1s \citep{Kollatschny11, Kollatschny13, Vietri18, Berton20a}. \par
In this young age scenario, the radio spectrum has multiple components. 
The low-frequency emission may be dominated by the strong star formation typical of NLS1s, while at higher frequencies the spectrum may rise again due to the misaligned relativistic jet component.
Albeit in general the emission from a relativistic jet would always be dominant over that from star formation at radio frequencies, this may not be the case in NLS1s, since their star formation rate is higher than in other AGN and it can contribute significantly to the radio emission \citep{Sani10, Caccianiga15}. \par
We remark that the first detection at 37~GHz occurred in 2015 \citep{Lahteenmaki18}. 
Assuming that the original peak was at 37~GHz and that the jet propagates between 0.3c and c, in 5 years the turn-off of the radio spectrum would have already moved between $\sim$32 to 25~GHz. 
Therefore, the convex feature should move toward lower energies, the detection rate at 37~GHz should decrease because the low-state flux density would decrease, and we should see the spectrum rising already at 9.0~GHz. 
New observations are therefore essential to test this scenario. 

\subsection{Ionized gas and free-free absorption}

As previously mentioned, in our sources the low-frequency (up to 10~GHz) emission is consistent with being produced by star formation processes, while the high frequency emission may be dominated by the relativistic jet.
There is another possible scenario that could account for their unusual behavior. 
Since we do not see a dominant jet component at JVLA frequencies, some kind of absorption must be active in the low-frequency regime. 
If we exclude synchrotron self-absorption, the other possible candidate may be instead free-free absorption. 
In the most extreme cases its spectral index may reach a value of $\alpha = -4$  \citep[e.g.,][]{Condon92, Falcke99b, Tingay03}.
In this case, assuming that the minimum of the spectrum is located at 10~GHz, and that this frequency represents the flux density observed in the low state of the jet, in the representative case of J1641+3454 we obtained a low-state flux density at 37~GHz of $\sim$103 mJy. 
This value is still below the detection threshold of Mets\"ahovi, as expected, and it requires that the source during flares increases its flux density by a factor of $\sim$4, which may be a rather common event. 
Furthermore, flares are usually associated with the ejection of new components in the relativistic jets that propagate downstream. 
These components usually have a higher Doppler factor with respect to the typical bulk motion in the jet ($3-\alpha_e$ with respect to $2-\alpha_e$, where $\alpha_e$ is the spectral index of the electron population), therefore at high frequencies their brightness may be strongly enhanced by relativistic effects.\par
In conclusion, free-free absorption may be a reasonable explanation for the spectrum observed in J1641+3454, but also in the other sources examined in this work, since their behavior is essentially identical. 
It is known that star formation in NLS1s is enhanced with respect to regular Seyfert galaxies \citep{Sani10}. 
Therefore, the circumnuclear gas could be ionized both by the AGN and by hot stars, and act as a screen for the jet at low frequencies. 
Another possibility is that the passage of the jet in the interstellar medium causes the formation of a cocoon of ionized gas \citep{Wagner11, Wagner12, Morganti15, Morganti17}, which could also be responsible for the free-free absorption \citep{Bicknell97}.
%The density required to have an optically thick gas in radio can be derived from the free-free absorption relation, using
%\begin{equation}
%    \nu_{\tau_\nu = 1} = \left[ \frac{1}{8.24\times10^{-2} T_e^{-1.35} N_e^2 d} \right]^{-\frac{1}{2.1}}
%\end{equation}
%where $\tau_\nu$ is the optical depth, $T_e$ is the electronic temperature, $N_e$ is the electron density, and d is the size of the ionized region. For typical values of $T_e = 10^4$K, and assuming $d=10$pc, a gas with density of $10^4$ cm$^{-3}$ would be optically thick for all the radiation below 15 GHz, which is close to what we observe.
In the JVLA bands, instead, the radio emission is dominated by the star forming activity happening farther from the nucleus, where the absorption becomes negligible. 
However, more detailed multiwavelength observations are needed to confirm this scenario, especially during the flaring events detected at 37~GHz. 

\subsection{Astrophysical implications}
As mentioned in the introduction, the radio-loudness parameter is still very often used as a sign of the presence of relativistic jets. 
However, our findings clearly show that such parameter is far from perfect, particularly among those sources that belong to the population A of the quasar main sequence.
These objects are typically characterized by a high Eddington ratio \citep{Boroson92}, which is usually associated with a dense circumnuclear environment \citep{Heckman14} and with enhanced star formation activity \citep{Chen09}.\par
The presence of these features can clearly affect the radio emission from population A sources. 
A high level of star formation significantly increases the radio luminosity of the sources, particularly at low frequencies (e.g., 5~GHz), and in some cases it can be so strong that the AGN may be classified as radio-loud even without a relativistic jet \citep{Caccianiga15, Ganci19}. 
On the other hand, the opposite situation can occur in sources like the ones discussed in this paper. 
If a screen of ionized gas is present around either the nuclear region or the relativistic jet itself, at low frequencies the radio emission will be suppressed, and a jetted object will be classified as radio-quiet (or radio-silent). 
In the case of a relatively faint relativistic jet produced by a low-mass black hole, radio-loudness can also be deceiving. 
As shown in the literature, radio-quiet or barely radio-loud sources can harbor relativistic jets, both on pc-scale \citep[e.g., Mrk 1239,][]{Doi15} and kpc-scale \citep[e.g., Mrk 783,][]{Congiu17}. 
Therefore, the use of 5~GHz flux density, or any other low frequency, to estimate radio-loudness can be strongly misleading. \par
This result can also have some consequences on the search for the parent population of $\gamma$-ray emitting NLS1s, that is how these aligned objects look like when they are observed at large angles \citep[for a review, see][]{Berton18c}. 
The less clear aspect of the parent population was if radio-quiet NLS1s had a role in it \citep[e.g., see][]{Berton16b, Berton18b}. 
Now, it seems that an unknown number of jetted NLS1s may be present within the radio-quiet/silent population.
When observed at small angles, these sources may be capable of producing the observed $\gamma$-ray emission, and they may therefore complete the picture of the parent population. \par
Finally, it is also worth noting that, if free-free absorption can be so effective in screening relativistic jets among population A objects, there may be a completely unknown population of radio-quiet objects capable of producing $\gamma$-ray emission, in analogy with what we see in J1641+3454. 
Such sources may even be associated with some of the unidentified $\gamma$-ray sources present in all the \textit{Fermi} catalogs. 
Given their lack of radio emission, their association to any nearby radio-quiet AGN would be far from obvious. \par
In conclusion, a physics-based classification, such as the jetted/non-jetted dichotomy \citep{Padovani17}, can provide a much more significant view on the nature of AGN. 
The radio-loudness parameters, instead, should be used only in extreme cases when nothing else is available, particularly among population A sources, and any result based on it should be taken with extreme care.

\section{Summary}
In this paper we present the results of new observations carried out with the Very Large Array of seven narrow-line Seyfert 1 galaxies (NLS1) formerly classified as radio-quiet or radio-silent. These NLS1s have been detected multiple times during flaring episodes at 37~GHz at Mets\"ahovi Radio Observatory, suggesting the presence of relativistic jets propagating from their core. However, our deep radio observations at frequencies between 1.6 and 9.0~GHz showed that these sources, in analogy with previous observations of radio surveys, are very faint or even not detected. Considering their low luminosity and brightness temperature, their radio spectrum below 9.0~GHz seems dominated by star forming activity, while the jet component is completely invisible. At high frequencies, instead, the jet must be present, otherwise the detection at Mets\"ahovi would be impossible. Therefore, their radio spectrum has a convex shape. While the low-frequency regime is dominated by a steep spectrum typical of star formation, we discuss a few possibilities to account for the extremely inverted spectrum observed at higher frequencies.
\begin{itemize}
\item \textit{Coronal origin}: coronal emission in radio-quiet quasars may appear as a turnover toward high frequencies, and be extremely variable. However, the flux densities measured at Mets\"ahovi are one order of magnitude larger than any previous emission attributed to the corona, therefore this solution is unlikely. 
\item \textit{Synchrotron self-absorption}: the inverted part of the spectrum may be the optically-thick emission from the relativistic jet, peaking at high frequencies as seen in extremely young radio galaxies such as high-frequency peakers. However, the kinematic age of the jet predicted by this model is rather small ($\sim$10 years), so we would not expect to see many objects of this kind. 
\item \textit{Free-free absorption}: NLS1s are well-known for their strong star formation, which along with the AGN activity can produce large quantities of ionized gas. The latter could be responsible for screening the jet emission at low frequencies, preventing us from seeing the small-scale jet present in these NLS1s. 
\end{itemize}
This result implies that radio-loudness is not a good parameter to identify the presence of relativistic jets in AGN, particularly in NLS1s and in similar sources. If our interpretation is correct, several relativistic jets could be hidden in AGN without any prominent radio emission. Additional observations at higher frequencies are indispensable to fully constrain the spectral behavior of these peculiar sources.

\begin{acknowledgements}
G.T. acknowledges partial support by the National Science Foundation under Award No. AST-1909796.
The National Radio Astronomy Observatory is a facility of the National Science Foundation operated under cooperative agreement by Associated Universities, Inc. 
This publication makes use of facilities and data obtained at the Mets\"ahovi Radio Observatory, operated by Aalto University, Finland. This research has made use of the NASA/IPAC Extragalactic Database (NED) which is operated by the Jet Propulsion Laboratory, California Institute of Technology, under contract with the National Aeronautics and  Space Admistration. 
\end{acknowledgements}

\bibliographystyle{aa}
\bibliography{./biblio}

\end{document}